\documentclass[runningheads]{svmult}

\usepackage{makeidx}   
\usepackage{graphicx}  
\usepackage{subeqnar}  
\usepackage{multicol}  
\usepackage{physprbb}  
\makeindex             

\def\BP{Ballesteros-Paredes}
\def\grad{{\bf \nabla}}
\def\gtsima{$\; \buildrel > \over \sim \;$}    
\def\gtrsim{\lower.5ex\hbox{\gtsima}}           
\def\ltsima{$\; \buildrel < \over \sim \;$}    
\def\lesssim{\lower.5ex\hbox{\ltsima}}           

\def\le{\lambda_{\rm eq}}
\def\lf{\lambda_{\rm F}}
\def\tc{\tau_{\rm c}}
\def\td{\tau_{\rm d}}
\def\tk{\tau_{\kappa}}
\def\ts{\tau_{\rm s}}
\def\tu{\tau_{\rm u}}
\def\u{{\bf u}}
\def\VS{V\'azquez-Semadeni}

\begin{document}
\title*{Thermal Instability and Magnetic Pressure in the Turbulent
Interstellar Medium} 
\titlerunning{Thermal Instability and Magnetic Pressure in the Turbulent ISM}
%
\author{Enrique V\'azquez-Semadeni\inst{1}\and Adriana Gazol\inst{1}\and
Thierry Passot\inst{2} \and Javier S\'anchez-Salcedo\inst{3}}
\authorrunning{V\'azquez-Semadeni et al.}
%
%
\institute{Instituto de Astronom\'\i a, UNAM, Campus Morelia,
Apdo. Postal 3-72, Morelia, Michoac\'an, MEXICO \and CNRS, Observatoire de la 
C\^ote d'azur, B.P. 4229, 06304, Nice, C\'edex 4, France \and
Instituto de Astronom\'\i a, UNAM, Apdo. Postal 70-264, M\'exico,
D.F., 04510, M\'exico }

\maketitle              

\begin{abstract}
We review recent results on the nonlinear development of thermal
instability (TI) in the context of the
turbulent atomic interstellar medium (ISM), in which correlated density and
velocity fluctuations are present, as well as forces other than the
thermal pressure gradient. First, we present a brief summary of the
linear theory, remarking that, in the atomic ISM, the condensation mode is
unstable but the wave 
mode is stable at small scales. Next, we revisit the growth of isolated
entropy perturbations in initially unstable gas, as a function of the
ratio of the cooling to the dynamical crossing times $\eta$.
The time for the dynamical transient state to subside ranges from 4 to
30 Myr for initial density perturbations of $20\%$ and sizes 3 to 75 pc. 
When $\eta \ll 1$, the condensation produces locally supersonic motions
and a shock propagates off the condensation, bringing the 
surrounding medium out of thermal equilibrium. 
Third, we consider the evolution of
{\it velocity} perturbations, maintained by a random forcing,
representing turbulent energy injection to the ISM from stellar
sources. These perturbations correspond to the wave mode, and are stable
at moderate amplitudes and small scales, as confirmed numerically. 

We then consider the behavior of magnetic pressure in turbulent
regimes. Various observational and numerical results suggest that the
magnetic pressure does not 
correlate well with density at low and intermediate densities. We
propose that this is a consequence of the slow and fast
modes of nonlinear MHD waves being characterized by different scalings of the
magnetic field strength versus density. This lack of correlation
suggests that, in fully turbulent regimes, the magnetic field may not be a
very efficient source of pressure, and that polytropic descriptions of
magnetic pressure are probably not adequate. 

Finally, we discuss simulations of the ISM (and resolution issues) 
tailored to investigate the
possible existence of significant amounts of gas in the ``lukewarm''
temperature range between the warm and cold stable phases. The
mass fraction in this range increases, and the phase segregation
decreases, as smaller scales are considered. We attribute this to two
facts: the enhanced stability of moderate, adiabatic-like velocity
fluctuations with $\eta \gg 1$ and the recycling of gas from the dense
to the diffuse phase by stellar energy injection.
Moreover, the magnetic field is not strongly turbulent there, possibly 
providing  additional stability. We
conclude by suggesting that the gas with unstable temperatures can be
observationally distinguished through simultaneous determination of two
of its thermodynamic variables.

\end{abstract}

\section{Introduction}\label{sec:intro}

The fact that the neutral atomic interstellar medium (ISM) is most
likely thermally bistable \cite{Pik68,FGH69,Wol95}
has had a great impact on our picture of interstellar
structure formation. Indeed, in two of the most influential models of
the ISM to date, the famous two- and three-phase models of the ISM
of Field, Goldsmith \& Habing \cite{FGH69} and McKee \& Ostriker
\cite{MO}, the concepts of 
thermal and pressure equilibria played a fundamental role, so that
distinct {\it phases} (thermodynamic regimes with different density and
temperature, but the same pressure) were predicted to coexist in
pressure equilibrium. These phases correspond to stable
thermal-equilibrium (i.e., heating-cooling balance) temperature regimes, and
are separated by unstable regimes that, in those models, were therefore
not expected to be present in the ISM. An opposite view was taken in the
so-called time-dependent model of the ISM of Gerola, Kafatos \&
McCray \cite{GKM74},
which was presented as an alternative to the pressure equilibrium
two-phase model, and which made radically different assumptions: a
constant density in the presence of stochastic, local heating events
that caused strong local fluctuations of pressure and temperature, because
the cooling and recombination times are comparable or shorter than the
time between successive exposures of a given gas parcel to those heating
events. This model predicted that significant amounts of gas should be
in the unstable range, as they cooled after the transient heating
events. More recently, Lioure \& Chi\'eze \cite{LC90} have considered
models with a continuous recycling of gas
among the various gas phases due to stellar energy injection,
also concluding that significant amounts 
of gas should populate the unstable temperature range in the
ISM. Note that the three-phase model \cite{MO} did consider the
existence of local fluctuations in the pressure, although it was still
based on the premise of ``rough pressure balance''.

Nevertheless, both the equilibrium and the time-dependent models omitted
a number of important aspects in the ISM budget. The multiphase
equilibrium models essentially neglected the possibility of large
pressure fluctuations in the ISM. The time dependent model instead
included this possibility as a fundamental premise, but neglected the
fact that such pressure fluctuations should induce motions, which 
should in general be turbulent (i.e., spanning a wide range of scales),
and in turn cause strong density fluctuations \cite{Elm93,BVS99}.
Moreover, both the time-dependent and the three-phase
models omitted other important agents of the ISM, such as magnetic fields,
rotation, and cosmic rays. Elmegreen \cite{Elm91,Elm94} performed a
combined instability 
analysis including self-gravity, cooling and heating, and magnetic
fields, but the effects of turbulence, which is an inherently nonlinear
phenomenon, can only be dealt with by means of numerical simulations of
the gas dynamics in the Galactic disk in the presence of thermal
instability (TI). The role of
the turbulent motions may be crucial. In fact, realistic cloud/intercloud
structure has been reported in models incorporating turbulence from
stellar-like driving and cooling, but not necessarily a thermally
bistable regime \cite{BL80,CP85,CB88,RB95,VPP95,PVP95,GP99}. Similar
results have been reported for pressureless (Burgers-like) models
with stellar driving \cite{SC99,CS01}, and for simulations of
interacting nonlinear MHD waves \cite{Elm97}. Thus, it is important to
investigate the role of TI in determining the distribution of the physical
variables (density, temperature, velocity) of the flow, and, in
particular, the degree to which phase segregation, as was proposed in
the multiphase models, is realized, in the context of a turbulent ISM
with multiple sources of
turbulent energy at a variety of scales, such as stellar winds, supernova
explosions, spiral arm passage, magnetorotational instabilities
\cite{SB99}, etc., besides TI.

Although the nonlinear development of TI has been studied extensively for 
decades now (e.g.,
\cite{Gol70,SMS72,Muf74,Muf75,OMB82,Sas88,DBS88,ML89,ML90,BMM90,Kri90,KSFR90,KLR00,BL00}),
only recent work has started to investigate the interplay between TI
and the turbulent nature of the ISM, such as, for example, the triggering of
TI by external compressions (\cite{HP99,HP00,KI00}) and the possibility
that the TI itself may contribute to the generation of turbulence in the
ISM \cite{WSK00,KI01,KN02}. 

However, the fact that additional energy sources feeding ISM turbulence
besides TI itself, such as stellar energy injection, or large-scale
gravitational or magnetic instabilities, has additional
implications. First, the very presence of
strong motions implies that transport (advection) should be important,
while traditionally conductive processes have received more attention
(e.g., \cite{ZP69,BM90,Mee96}). Second, these transport processes may
imply the existence of constant fractions of gas transiting through the
unstable regime, and erase, to some extent, the phase segregation
expected in multiphase models. These expectations are furthered by several
observational studies (e.g., \cite{DST77,KSG85,SF95,FS97,Hei01}) that have 
suggested that the fraction of gas in the unstable range between the
cold and warm phases of the atomic ISM is substantial. 

Another important issue to consider is the fact that the ISM is
magnetized, which suggests the possibility that turbulent magnetic
pressure may supplement thermal pressure and somehow counteract TI. The
condensation process in a magnetized medium has been studied by several
workers (see, e.g. \cite{Fie65,Elm97,OMB82,Loe90,HP00}; see also the
references given in \cite{HP00}), concluding that,
although condensation can be inhibited under some circumstances, it is
in general possible. However, those studies have not considered the case 
of TI developing in an externally-driven turbulent medium, except for
\cite{HP00}.

In this paper we review recent work and present new results concerning
the interplay between TI and turbulence in the warm and cool ISM. First,
we review the main aspects of the instability in \S
\ref{sec:quali}. Then, in \S \ref{sec:pure_TI}, we revisit the growth of 
isolated density fluctuations, focusing in particular on the late stages 
and the state of the gas surrounding the condensation. 
Next we discuss the development or suppression 
of growth in the presence of random velocity fluctuations, stressing that these
probably constitute the most common way of inducing density fluctuations in
the ISM. We then briefly discuss the nature of magnetic pressure 
on turbulent media (\S \ref{sec:magn_pres}) and its dependence on
density. In \S \ref{sec:full_ISM}, we discuss
the role of TI in numerical models of the ISM in the presence
of magnetic fields, the Coriolis force, modeled star formation,
self-gravity and TI, aiming at determining the fraction of unstable
gas, and at interpreting the results in the light of the previous
sections. We include an extensive discussion of numerical tests to
maximize the reliability of the results. Finally, we summarize
the results and mention a number of implications in \S
\ref{sec:conclusions}.

\section{Review of the linear theory and the role of the ratio of
cooling time to crossing time} \label{sec:quali}

The ISM in general is subject to heating and cooling processes that
cause it to have an effective thermal behavior very different from that
of an ideal gas. Our current understanding of the physical processes
responsible for the heating and cooling of the neutral atomic ISM can be 
found in Wolfire et al.\ \cite{Wol95}. Their effect is customarily
collected in
a net cooling function per unit mass ${\cal L} = \rho \Lambda - \Gamma$,
where $\rho$ is the gas mass density, $\Lambda$ is the cooling rate, and 
$\Gamma$ is the heating rate. In this paper we will use a piecewise
power-law fit to the ``standard'' cooling curve of \cite{Wol95}, of the
form 
\begin{equation}
\Lambda=C_{i,i+1} T^{\beta}_{i,i+1}  \;\;\;\; {\rm for} \;\;\; T_{i}\leq
T< T_{i+1}, \label{eq:cooling}
\end{equation}
with the coefficients and exponents given in Table \ref{tab:cooling},
and a constant heating rate $\Gamma=\Gamma_0$, which is a reasonable
approximation to the weak dependence $\Gamma \sim \rho^{0.2}$ across the
density range of the atomic medium \cite{Wol95}. The fit is obtained by
first fitting a piecewise power law to the standard thermal
equilibrium-pressure vs.\ density curve of \cite{Wol95} to obtain the
exponents of the cooling curve, and then determining the coefficients by
equating the cooling rate to the constant heating $\Gamma_0=0.015$ erg
s$^{-1}$g$^{-1}$. The values of the coefficients have units of 
erg s$^{-1}$g$^{-2}$cm$^{3}$K$^{-\beta_{i,i+1}}$.

\begin{table}
\caption{Cooling function parameters}
\begin{center}
\renewcommand{\arraystretch}{1.4}
\setlength\tabcolsep{5pt}
\begin{tabular}{cccccccc}
\hline\noalign{\smallskip}
Interval  & $T_i/$K	& $n_i/$cm$^{-3}$	& $C_{i,i+1}^{\mathrm a}$& $\beta_{i,i+1}$ \\
\hline\noalign{\smallskip}
		& 15.0	& 80.0	\\
(1,2)		&	&		& $3.42 \times 10^{16}$	& 2.13 \\
		& 141.	& 7.00	\\
(2,3)		&	&		& $9.10\times 10^{18}$	& 1.00 \\
		& 313.	& 3.16	\\
(3,4)		&	&		& $1.11 \times 10^{20}$	& 0.565 \\
		& 6101. & 0.59	\\
(4,5)		&	&		& $2.00 \times 10^{8}$	& 3.67	\\
		& $10^5$& 	\\
\hline
\end{tabular}
\end{center}
$^{\mathrm a}$ In units of erg s$^{-1}$g$^{-2}$cm$^{3}$K$^{-\beta_{i,i+1}}$.
\label{tab:cooling}
\end{table}

Subject to these processes, and in the
absence of magnetic fields, self-gravity and other agents, the dynamics
of the gas are described by the equations \cite{Fie65,Mee96}
\begin{eqnarray}
\frac{d \rho}{dt} + \rho \grad \cdot \u = 0, \label{eq:cont}\\
\rho \frac{d \u}{dt} = -  \grad P \label{eq:mom}\\
\frac{1}{\gamma-1} \frac{dP}{dt} + \frac{\gamma}{\gamma -1} P \grad
\cdot \u + \rho {\cal L(\rho,T)} - \grad \cdot (K \grad T) =0,
\label{eq:int_en} 
\end{eqnarray}
where \u\ is the fluid velocity, $P$ is
the thermal pressure, $T$ is the temperature, $\gamma$ is the heat
capacity ratio, $K=K(T)$ is the thermal conductivity, and the ideal-gas
equation of state, $P=\rho R T/\mu$, is assumed, with $R$ being the
universal gas constant and $\mu$ the mean molecular weight in units of
the hydrogen mass. For clarity, we remind the reader that 
$R=k_{\rm B}/m_{\rm H}$, where $k_{\rm B}$ is the Boltzmann constant, and
$m_{\rm H}$ is the mass of the Hydrogen atom.

The linear analysis leading to TI was first performed in full in the
classic paper of Field \cite{Fie65}, and later generalized to the case
of a flow in motion \cite{Hun70,Hun71}. Other useful, more recent
presentations may be found in \cite{Elm91b,Shu92,Mee96}.
The analysis, assuming perturbations proportional to $\exp(n t + i
{\bf k} \cdot {\bf x})$ in all variables, yields the cubic dispersion 
relation
\begin{equation}
n^3 + n^2 \left[\frac{N_\rho}{c_V} + \frac{c k^2}{k_K} \right] + n  c^2
k^2 + c^2 k^2 \left[\frac{N_P}{c_P} + \frac{c k^2}{\gamma k_K} \right] = 0,
\label{eq:disp_rel}
\end{equation}
where $n$ is the growth rate, $k$ is the wavenumber,
$c\equiv \gamma kT/\mu$ is the adiabatic sound speed, and
\begin{equation}
N_\rho \equiv \left(\frac{\partial \cal L}{\partial
T}\right)_\rho, ~~~~~ N_P \equiv \left(\frac{\partial \cal 
L}{\partial T}\right)_P = \left[N_\rho
- \frac{\rho_0}{T_0} \left(\frac{\partial\cal
L}{\partial \rho}\right)_T\right], \label{eq:Ns}
\end{equation}
and 
\begin{equation}
k_K \equiv \frac{R}{\mu} \frac{c \rho_0}{(\gamma-1) K}. \label{eq:k_K}
\end{equation}
As pointed out by Field, $k_K$ is the mean free path of the gas
molecules (see also \cite{Shu92}, p.35). 

The dispersion relation (\ref{eq:disp_rel}) has three roots, one of them
being always real, and the other two being either a complex conjugate
pair, or a pair of real numbers. There is instability whenever
$\Re(n)>0$, where $\Re()$ denotes the real part. The positive real root 
corresponds to exponential growth without propagation, and is thus
called the {\it condensation} mode. The pair of complex roots corresponds
to oscillatory behavior, and is thus called the {\it wave mode}. This
mode grows in amplitude (i.e., is {\it overstable}) when the real part
of those roots is positive. This nomenclature 
was extended by Field to the case 
when the roots are real, in which case he said the wave mode is {\it
overdamped} (i.e., does not oscillate). 

The condensation mode is unstable if the so called ``isobaric''
criterion, namely
\begin{equation}
N_P < 0, \label{eq:isob_crit}
\end{equation}
is satisfied. However, when only this criterion is satisfied, the growth 
rate vanishes as $k \rightarrow 0$. For the growth rate to remain finite 
at long wavelengths, it is necessary to also satisfy the ``isochoric''
criterion, which reads
\begin{equation}
N_\rho < 0. \label{eq:isoch_crit}
\end{equation}

The corresponding instability criterion for the wave mode is the so
called ``isentropic'' (or ``adiabatic'') criterion, reading
\begin{equation}
\frac{N_P}{c_P} - \frac{N_\rho}{c_V} > 0, \label{eq:isent_crit}
\end{equation}
where $c_P$ and $c_V$ are the specific heats at constant pressure and at 
constant volume, respectively. Note that $c_V = R/\left[(\gamma -1) \mu
\right]$. For the cooling and heating functions adopted
here, the isobaric, isochoric and isentropic criteria respectively imply 
$\beta < 1$, $\beta < 0$ and $\beta < 1/(1-\gamma) =-3/2$.

Since in the atomic ISM only the isobaric 
criterion is satisfied, and given the focus of this paper on that
medium, we will concentrate on this case in what follows. It is worth
recalling that isobaric and isochoric perturbations
(accomplished, for example, by setting up density fluctuations at
constant pressure or temperature fluctuations at constant density,
respectively), are generically referred to as {\it entropy} preturbations, 
since they imply a variation of the ratio $P/\rho^\gamma$, which remains
constant for reversible isentropic processes. We now discuss the
evolution of entropy perturbations in detail.

\subsection{Entropy perturbations} \label{sec:entropy_pert}

\begin{figure}[b]
\begin{center}
\includegraphics[width=.5\textwidth]{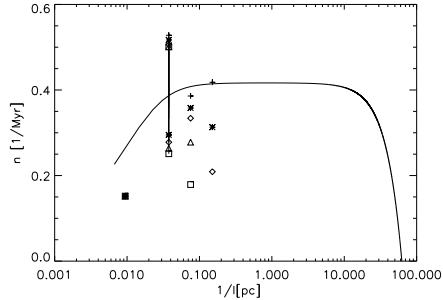}
\end{center}
\caption[]{Growth rate as a function of wavenumber for our piecewise
power-law fit to the ``standard'' cooling function of Wolfire et al.\
\cite{Wol95} and a constant heating rate. The wavenumber at which the
growth rate becomes zero corresponds to the Field length. The points
show the numerical growth rates in simulations of the linear growth of
2.5\% isobaric density perturbations in
boxes of physical size 150 pc (see \S \ref{sec:numerical}) for several
values of the mass diffusion coefficient: 
$\mu=0$ ({\it crosses}); $\mu=0.001$ ({\it asterisks}); $\mu=0.002$ ({\it
diamonds}); $\mu=0.004$ ({\it triangles}); $\mu=0.008$ ({\it
squares}). For perturbations with $l=1/0.05=18.75$ pc (1/8 of the box
size), which are in the transition regime between long and short
wavelengths, growth rates are shown that have either zero (appropriate
for the short-wavelength regime; points
lying below the theoretical curve) or self-consistent velocity and
pressure fluctuations (appropriate for the long-wavelength regime;
points lying above the curve).
} 
\label{fig:wvsk}
\end{figure}

Figure \ref{fig:wvsk} shows the growth rate as a function of wavenumber
for our fit to the ``standard'' cooling curve of \cite{Wol95}, eq.\
(\ref{eq:cooling}) and a realistic value of the conductivity $K=K_0=10^{-6}
T_0^{5/2}$ erg cm$^{-1}$ s$^{-1}$ K$^{-1}$ \cite{All00}, with $T_0=2400$
K. In the case of the pure 
development of the instability, without any external forcing processes,
there are three clearly distinct scale ranges for the wavelength
$\lambda$ that arise from the
presence of three characteristic time scales \cite{Mee96}: the dynamical
time, $\td$, in this case given
by the sound crossing time, $\ts = c/\lambda$; the cooling time, 
which for isobaric processes is given by\footnote{This expression is
obtained by linearizing the energy equation (\ref{eq:int_en}) and then
computing $\tc = \delta e/(d \delta e/dt)$ for an isobaric process.}
\begin{equation}
\tc = \frac{\gamma R}{(\gamma-1)\mu |N_P|}, \label{eq:growth_t}
\end{equation}
but which, to order of magnitude, is in general $\tc \approx c_V
T/(\rho \Lambda)$; and the conductive time, $\tk = \lambda^2/\kappa_0$,
where $\kappa_0 = K_0/\rho_0$ is the thermal diffusivity.
These time scales then define two characteristic length scales. First,
the so called Field length \cite{BM90} $\lf = 2 \pi/k_{\rm F}$, with
$k_{\rm F} = (|N_P|/\kappa_0)^{1/2}$, at which
$\tk \sim \tc$, and below which the growth of perturbations
is inhibited by thermal conduction. Second, the scale at which $\td \sim
\tc$, which we shall denote $\le$. In the remainder of the paper we
shall use the notation $\eta \equiv \tc/\td$. The three scale ranges are 
then the ``long'' ($\lambda \gg \le$), ``intermediate'' ($\le \gg
\lambda \gg \lf$) and ``short'' ($\lf \gtrsim \lambda$), where $\lambda
= 2 \pi/k$ is the perturbation wavelength. For the long-wavelength
range, $\eta \ll 1$, while for the intermediate- and short-wavelength
cases, $\eta \gg 1$. In what follows we will often discuss in terms of
$\eta$, as we consider that the relevant physical quantities involved are
the cooling and sound-crossing time scales, even though it is customary
in the literature to base the discussion on the perturbation wavelength.

In the case of short- and intermediate-wavelength perturbations
\footnote{For convenience, in the remainder of this paper, we will
group intermediate- and small-scale 
perturbations into the small-wavelength ($\eta >1$) category.}, the
condensation mode evolves nearly isobarically, as $\tc \gg \td$ and
thus sound waves have ample time to restore pressure equilibrium while
the gas cools. This also has the consequence that,
in the case of zero diffusivity, the growth rate asymptotically
approaches the cooling rate in the limit of large
wavenumbers. In the presence of diffusivity, the growth rate 
decreases again in the short-wavelength range, due to the action of
thermal conduction. These properties are illustrated in fig.\ \ref{fig:wvsk}.

\begin{figure}[b]
\begin{center}
\includegraphics[width=.8\textwidth]{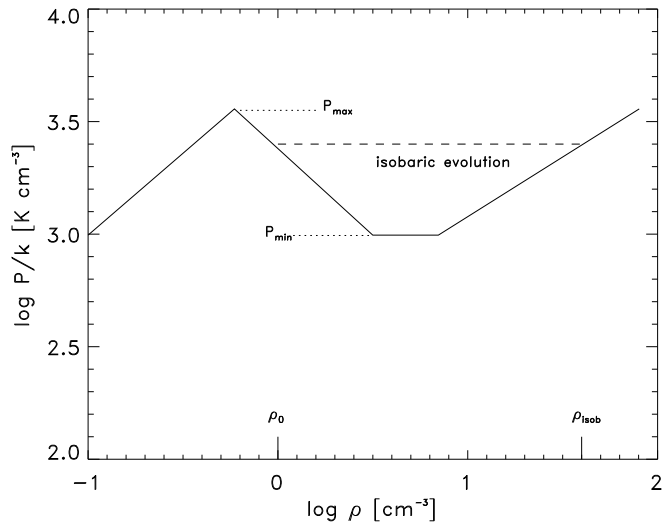}
\end{center}
\caption[]{Thermal-equilibrium piecewise polytropic behavior of the
pressure for the cooling functions used in Papers I and IV. $\rho_{\rm
isob}$ is the value of the density within the cold phase whose
corresponding thermal-equilibrium pressure equals that of the mean
density $\rho_0$.}
\label{fig:Peq}
\end{figure}

In the opposite limit of very large wavelengths, the condensation
mode initially behaves isochorically (even though the isochoric
criterion is not satisfied), since $\eta \ll 1$, so that the
cooling acts much more 
rapidly than the sound waves can travel accross the perturbation to
restore pressure balance. This implies that large pressure gradients can 
be set up, which in turn can trigger strong motions that can become
locally supersonic \cite{Fie65,MS87,Bal95,Mee96}. Moreover, in this regime,
the thermal pressure is given by the condition of
thermal equilibrium, because the rapid cooling always allows its
establishment. Figure \ref{fig:Peq} shows the equilibrium-pressure
versus density for our piecewise cooling function
(\ref{eq:cooling}). The slope of this graph constitutes an effective
polytropic exponent given by $\gamma_{i,i+1} = 1-1/\beta_{i,i+1}$ (c.f.\ 
eq.\ [\ref{eq:cooling}]), so that the pressure behaves as $P \propto
\rho^{\gamma_{i,i+1}}$. In this figure, we
denote by $\rho_{\rm isob}$ the density value within the cold phase
that corresponds to the same pressure as that at the mean density.
The instability under the isobaric mode is seen as the negative-slope
range 0.6 cm$^{-3} \lesssim \rho \lesssim 3.2$ cm$^{-3}$ (equivalent in
thermal equilibrium to the temperature range $300 \lesssim T \lesssim
6000$ K). Note also the marginally stable, $\gamma_{\rm
eff} = 0$ behavior in the range 3.2 cm$^{-3} \lesssim \rho \lesssim
7.1$ cm$^{-3}$. 


In this same limit (long wavelengths), the growth rate asymptotically
approaches \cite{Mee96} 
\begin{equation}
n_{\rm long} = \pm k c \left(-\frac{N_P}{N_\rho}\right)^{1/2},
\label{eq:rate_long}
\end{equation}
which is of the order of the inverse of the sound crossing time,
explaining its vanishing as $k \rightarrow 0$ (fig.\
\ref{fig:wvsk}). Note that the situation is different when the 
isochoric criterion is satisfied, in which case, the growth rate remains 
roughly constant at $\tc^{-1}$ over the long- and
intermediate-wavelength ranges.


\subsection{Adiabatic perturbations} \label{sec:adiab_pert}

The case of adiabatic perturbations is perhaps the
most relevant for a turbulent ISM since the adiabatic
condition implies that the density and pressure gradients have the same
sign, a situation which is naturally accomplished by means of a
compressive motion acting on time scales much shorter than the cooling
time. In contrast, for entropy perturbations the density and pressure
gradients can have opposite signs. We refer to the latter case as a
``reversed'' 
pressure gradient, while we say that a pressure gradient with the same
sign as the density gradient is ``regular''.

In the presence of velocity fluctuations, a new characteristic time
scale appears in the system, namely the bulk-velocity crossing
time, $\tu =\lambda /u$, where $u$ is the characteristic velocity of the
perturbation. Thus, it is convenient to redefine the dynamical time scale
as $\td = \min(\ts,\tu)$, so that the definition $\eta = \tc/\td$ can be
preserved. Adiabatic perturbations with $\eta \gg 1$ become
unstable when the adiabatic instability criterion (\ref{eq:isent_crit})
is satisfied \cite{Fie65,Shu92}, and excite the wave mode of the
instability, which in this case consists of nearly adiabatic sound waves
with growth rate $n = ikc +(1/2) (N_P/c_P-N_\rho/c_V)$, where the
imaginary 
part gives the propagation speed and the real part gives a modulation
that grows only if the adiabatic criterion is satisfied
(overstability). However, at $\eta <1$ (long wavelengths), the complex 
conjugate roots become real, and the wave mode becomes a condensation
mode, which is unstable whenever $N_P/(\gamma N_\rho)<0$
\cite{Shu92}. For $N_\rho >0$, as is usually the case, this reduces to
the isobaric criterion. Thus, in
the atomic ISM, {\it wave-like (or ``velocity'') perturbations are linearly
unstable only for
$\eta <1$, i.e, in the long-wavelength limit}. For  nonlinear perturbations,
however, the density increase in the waves can accelerate the cooling
and locally cause $\eta$ to become $<1$, allowing the instability to
proceed even in cases of perturbations of initial short-wavelength
perturbations. A similar effect can occur even when the compression acts 
on the warm stable phase \cite{HP99,KI00}. We will discuss this further
in \S \ref{sec:vel_fluc}.

\subsection{Entropy vs.\ adiabatic fluctuations}
\label{sec:ent_vs_adiab}

 It is now important to ask what kind of
processes in the real world can generate the two kinds of fluctuations:
entropy or adiabatic. In the absence of initial fluid motions, entropy 
fluctuations can only be produced by locally varying the
cooling-to-heating ratio. On the other hand, if velocity 
fluctuations are used as the driver of the density fluctuations, as
occurs in a turbulent medium, then the density fluctuations can either
behave as entropy or as adiabatic fluctuations 
depending on scale and on the velocity amplitude. The production of
entropy-like perturbations clearly requires that $\tu \gg \tc$, so that
the thermal response of the flow proceeds under thermal equilibrium. At
small scales, where $\eta \gg 1$, this then implies that $\tu \gg \tc
\gg \ts$, so that the motions have to be essentially quasi-static. At
large scales, where $\eta \ll 1$, we see that even supersonic motions
can produce entropy-like fluctuations, as long as $\tu, \ts \gg
\tc$. The remaining possibility, i.e, $\tu \sim \ts \ll \tc$, occurring
for finite-amplitude velocity fluctuations at small scales, causes
adiabatic-like perturbations.

Another important distinction is that,
for entropy perturbations, the motions are driven by the
thermal pressure gradient generated by the instability, and tend to
restore pressure equilibrium. Instead, in the case of externally-driven
velocity perturbations, the motions drive the density and pressure
gradients, which then feed back on the cooling and the motions
themselves. Thus, the cause-effect relationship between the motions and
the thermal pressure gradient are essentially reversed in the two cases.

\subsection{The magnetic case} \label{sec:MHD_TI}

The linear instability analysis in the presence of a uniform magnetic
field {\rm B} was also studied by Field \cite{Fie65}. Here we just briefly
summarize his main results, and then discuss some recent work in the
nonlinear regime. 

Qualitatively, Field concluded that the inclusion of the magnetic field
should introduce three main modifications to the non-magnetic
results. First, the wave mode splits into three modes, which correspond
to the three modes of MHD waves: Alfv\'en, slow and fast. Of these, the
Alfv\'en mode is ``neutral'', in the sense that it does not interact
with the instability, since it is strictly non-compressive, while the
fast and slow modes are governed by the same isentropic criterion. Second, 
the condensation mode is unaffected when the vector wavenumber {\bf k} is
parallel to {\bf B}, while the field has a stabilizing effect when {\bf
k} is perpendicular to {\bf B}. Finally, heat conduction is greatly
reduced in the direction perpendicular to the field because of the
spiraling of electrons between collisions. In summary, no
major modifications to the overall picture were foreseen by Field, 
even though the dispersion relation changes from cubic to
fifth-degree. These results were verified numerically by Goldsmith
\cite{Gol70}. 

More recently, Loewenstein \cite{Loe90} has extended Field's linear
analysis to the  
case of a stratified backgound medium, showing that condensation modes
do exist in cooling flows, and that, over a certain wavenumber range,
the presence of the magnetic field suppresses the damping of the
instability due to conduction. In work more closely related to our focus in
this paper, Hennebelle \& P\'erault \cite{HP00} have investigated the 
role of a strong compression wave, interpreted as a turbulent velocity
fluctuation, on triggering the {\it nonlinear} instability in the
linearly stable diffuse phase in a magnetized medium already
segregated into two phases. These authors showed that the instability
can indeed  be triggered when the directions of the compression and of
the initial 
magnetic field are oblique, giving the threshold values of the angle
between them for condensation to occur at various values of the
compression Mach number and of the magnetic field strength. 
In the forthcoming sections we will first describe results concerning 
the development of TI from an initially unstable medium under both
quiescent and turbulent conditions, in the non-magnetic case. Next we
discuss the nature of magnetic pressure in turbulent media, and finally
we will consider more complete models of the ISM incorporating all of
these agents and processes.

\section{Nonlinear evolution of entropy perturbations}\label{sec:pure_TI}

As a first step in our discussion of dynamical aspects of the
development of TI, in this section we discuss the conditions necessary
for the development of large velocities (possibly supersonic) and shocks
during the spontaneous (i.e., in the absence of external triggers)
condensation process of entropy perturbations in the unstable atomic
ISM, as a function of the parameter $\eta$. A detailed review of
the nature of the shocks has been presented in \cite{Mee96}. 
We are also interested in the duration of the dynamic
phase. Studies dealing with the development of supersonic motions and 
shocks as a consequence of the condensation process from the unstable
regime have mostly focused on the regimes of proto-galaxy-cluster,
proto-galactic, and proto-globular cluster clouds (e.g.,
\cite{DBS88,Sas88,ML89,ML90,BMM90,Kri90,KLR00}). In particular, Sasorov
\cite{Sas88} pointed out that the development of TI in three dimensions
should give rise to flattened structures, similar to the ``pancakes''
formed by gravitational contraction in the cosmological flow.

In the context of the ISM, the nonlinear development of isolated entropy
fluctuations was initially studied by Goldsmith \cite{Gol70} and
Schwartz, McCray \& Stein \cite{SMS72}, who found that the condensation
of small-scale ($\eta >1$) perturbations occurred on time scales of
$\sim 1$--10 Myr, and produced clouds of densities $100 \times$ larger than
their initial values. Goldsmith also considered the case of large-scale
perturbations, finding the development of transonic velocities, and that 
the time required for reaching a true steady state is much longer than
the time of ``initial collapse'', not being reached by any of his
simulations. However,
these works were performed at very low resolutions, and did not discuss
the state of the surrounding gas in much detail.

More recently, Burkert \& Lin \cite{BL00} have
considered a cooling-only medium (i.e., without background 
heating), and suggested that a special clump scale can be selected by
the following mechanism. In the case of a globally cooling medium which
eventually exits the thermally unstable range in roughly one cooling
time, large-scale ($\eta <1$) fluctuations that cool isochorically do
not change their density appreciably before exiting the unstable range,
so that, after they do, the pressure gradient becomes regular again, and
the perturbation is erased. On the other hand, small-scale fluctuations
can eventually reach the regime of isochoric cooling with
$\eta<1$ as their density and local cooling rate increase, and then stop 
growing, except if they reach this transition stage already with
nonlinear amplitudes, in which case advection overtakes the pressure
gradient in promoting the compression, which then proceeds at an
accelerated rate. Thus, Burkert \& Lin suggested that the latter
fluctuations are the ones that dominate the fragmentation of a large
cloud into clumps, determining the clump properties.

In the presence of both cooling and heating,
S\'anchez-Salcedo et al.\ \cite{SVG02} (hereafter Paper I) have
recently investigated the evolution of perturbations as a function of
size (or, equivalently, $\eta$), focusing on the magnitudes of the
velocities that develop, the time scales for reaching a relatively
quiescent stage, and the state of the gas surrounding the condensation
after the latter has reached a quasi-stationary 
state. To this end, Paper I performed one-dimensional (1D),
high-resolution numerical simulations of the evolution of
Gaussian-shaped perturbations in the absence of any other physical
processes. The simulations use up to 7000 grid 
points, and non-uniform grid spacing, with central grid point density
$\sim 100\times$ that at the edges, in order to maximize
the resolution at the condensation center. The simulations solve the gas
dynamic equations in the presence of heating and cooling parameterized
as in eq.\ (\ref{eq:cooling}), reading
\begin{eqnarray}
\frac{D \ln \rho}{Dt} = - \frac{\partial u}{\partial x} + \frac{1}{\rho}
\mu \frac{\partial^2 \rho}{\partial x^2}, \label{eq:1Dcont}\\
\frac{D \u}{Dt} = -\frac{1}{\rho} \frac{\partial P}{\partial x} + f +
\frac{1}{\rho} \frac{\partial (2 \nu \rho S)}{\partial x},  \label{eq:1Dmom}\\
T \frac{Ds}{Dt} = \Gamma - \rho \Lambda + 2 \nu S^2
+\hbox{diffusion term}, \label{eq:1Dent}
\end{eqnarray}
where $D/Dt=\partial /\partial t + u \partial/ \partial x$ is the
Lagrangian derivative, $S = (2/3)  \partial u /\partial x$
is the generalized strain tensor in the 1D case, $s$ is the entropy per
unit mass, $f$ is the random forcing (described and used in \S
\ref{sec:vel_fluc}), and the other quantities have their usual
meanings. A shock-capturing viscosity of the form
\begin{equation}
\nu_s = \nu_0 + c_\nu \delta x^2 \max(0,-\nabla \cdot \u)
\label{eq:shock_visc}
\end{equation}
is used, with $c_\nu$ a constant.
The last term in eq.\ (\ref{eq:1Dcont}) is an artificial mass diffusion
term, necessary for smoothing excessively large density gradients. 
The term $2 \nu S^2$ in eq.\ (\ref{eq:1Dent}) is the viscous heating of
the gas. The diffusion term in the same equation has the same form
as the mass diffusion term, and is included to guarantee that any mass
redistribution occurs with the corresponding entropy redistribution.
The coefficients $\mu$ of these terms are maintained at very low values
so that these diffusivities are comparable to the numerical
diffusivity, giving a diffusive scale of size 2--3 grid zones. All the
results reported in this section have been 
subjected to convergence tests to guarantee that they are not altered by 
changing the resolution (see Paper I).

\begin{figure}[b]
\begin{center}
\includegraphics[width=1.\textwidth]{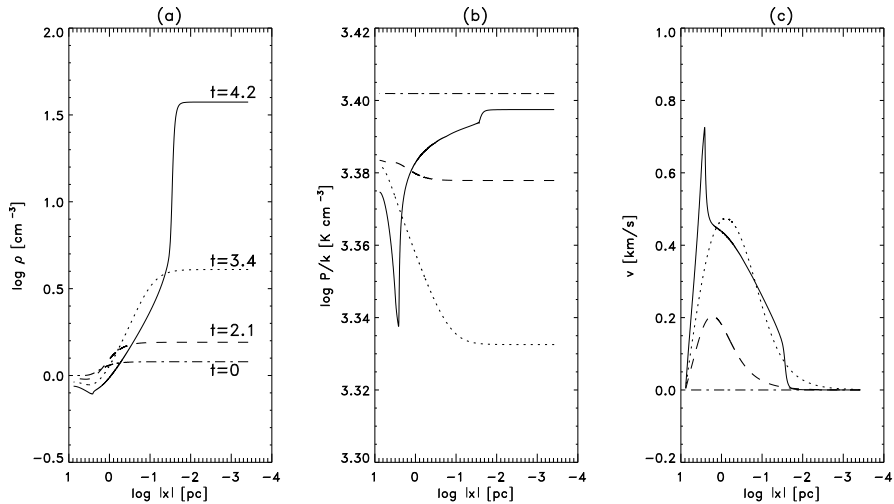}
\end{center}
\caption[]{Time development of run DEN3. The density,
pressure and velocity profiles at different times are plotted in panels 
(a), (b) and (c), respectively. The times corresponding to each line
type are indicated in frame (a) in Myr, and the same line labeling is
used in frames (b) and (c).}
\label{fig:den3}
\end{figure}

The simulations start at a density of 1 cm$^{-3}$, roughly the mean ISM
density in our galaxy, as pointed out in \S \ref{sec:MHD_TI} from
\cite{Fer01}, which lies in the unstable range of the cooling curve. The 
equilibrium temperature at this density is $T \approx 2400$ K. The
scale $\le$ at which the cooling and sound crossing times are equal is
$\le \sim 10$ pc, while the Field length under these conditions is $\lf \sim 
0.015$ pc (cf.\ \S \ref{sec:entropy_pert} and fig.\ \ref{fig:wvsk}).

\begin{figure}[b]
\begin{center}
\includegraphics[width=1.\textwidth]{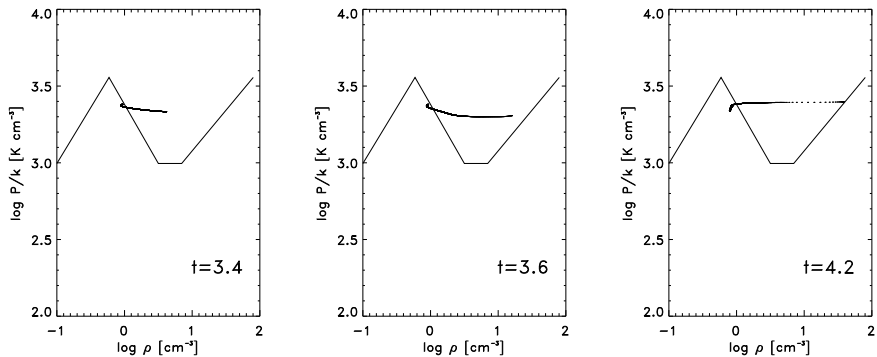}
\end{center}
\caption[]{Time development of run DEN3 in $P$--$\rho$ phase space, at
the times indicated in each frame. At $t=4.2$ Myr, although the cloud
has already formed, a substantial fraction of the points in the
simulation are still traversing the unstable range, albeit in a nearly
isobaric regime.}
\label{fig:den3_Prho}
\end{figure}

We first consider the case of small-scale ($\eta > 1$) entropy
perturbations, as they 
constitute the paradigm of cloudlet (i.e., small clouds of sizes 
$\lesssim 1$ pc and densities $\sim 50$ cm$^{-3}$) formation by TI in
the ISM (e.g., \cite{FGH69,Gol70,SMS72,Par87,LC90}). To this end, we
have performed a simulation, labeled DEN3, of the evolution of a
Gaussian density perturbation of 
20\% amplitude and a full width at half maximum (FWHM) of 3 pc. Figure
\ref{fig:den3} shows the density, pressure and  
velocity profiles of the cloud at various times until the time when
a ``cloud'' has formed and the accretion process has mostly subsided. 
Note the logarithmic $x$-axis, where $x$ is the distance to the
center of the cloud.
It is seen that the evolution is indeed quasi-isobaric, with
variations in the pressure of less than 2\%, and local Mach numbers which do
not exceed 0.2. By $t=4.2$ Myr, the condensation has essentially
completed its evolution, and reached the pressure-equilibrium density,
$\rho_{\rm isob}$. Figure \ref{fig:den3_Prho} shows the evolution of this run
on the $P$-$\rho$ diagram. The quasi-isobaric nature of the condensation
is clearly seen, especially at the final time, at which all points 
with densities higher than the mean have almost exactly the same
pressure. 

Note, however, that in figs.\ \ref{fig:den3} and \ref{fig:den3_Prho} a
population of points is 
still seen to continue flowing onto the condensation, and as it does,
it necessarily remains in the ``unstable'' range. In fact, the mass of this
gas amounts to $\sim 8$ times the mass in the condensation (counting
only the gas reached by the rarefaction wave). In fig.\
\ref{fig:den3}, it is seen that this occurs in the   
region $0 \gtrsim \log |x/({\rm pc})| \gtrsim -1.5$. 
The accretion and evacuation of the unstable gas will take very long
times to complete, because the reservoir of unstable gas outside
the cloud is very large ($\sim 90\%$ of the total mass), in agreement
with the remark by Goldsmith \cite{Gol70}. 
Moreover, note that {\it the inflowing gas is not truly
unstable}, as it does not lie on the equilibrium curve anymore. Instead,
at $t=4.2$ Myr, the density and pressure gradients have the same sign
throughout this region. Thus this gas does not have a tendency to
fragment any further, even though its density is in the ``unstable''
range. It can be said that this gas is {\it flowing} because of TI, but
once it is doing so it has no tendency to fragment any further. 

Finally, note that the
cloud formation time is not very short, and is significantly 
sensitive to the initial amplitude. Simulations with an initial
amplitude of 10\% require $\sim 5.5$ Myr to complete the
condensation. This time is comparable to the mean time between
successive exposures to passing shock fronts from supernova remnants and 
superbubbles \cite{KS00}, so that the condensations may have their
growth interrupted by external perturbations, as is the case in \S
\ref{sec:vel_fluc}.

\begin{figure}
\begin{center}
\includegraphics[width=.8\textwidth]{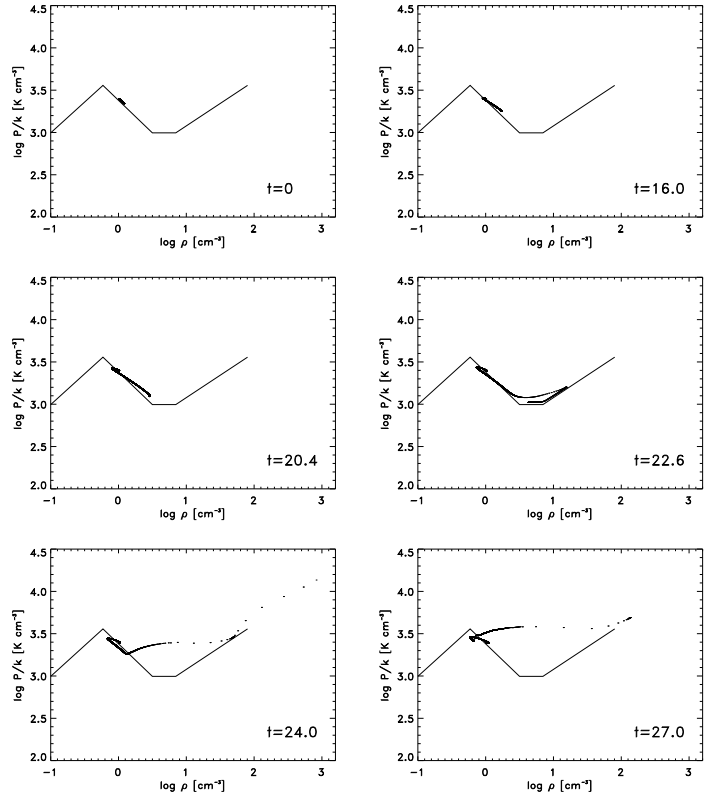}
\end{center}
\caption[]{Time development of run DEN75 in $P$--$\rho$ phase
space. Before the shock formation, the evolution proceeds along the
thermal equilibrium curve. Subsequently, the
outwards-propagating shock brings the outside medium out of thermal
equilibrium, and restores nearly pressure balance. Thus, the infalling
gas is traversing the density ``unstable'' range, but in nearly isobaric 
(inertial) conditions, rather than along the thermal equilibrium curve.}
\label{fig:den75_Prho}
\end{figure}

Let us now consider the opposite case of a large-scale entropy perturbation,
with FWHM=75 pc and $\eta \sim 0.04$, in a box of 250 pc. We refer to
this simulation as run 
DEN75.
\begin{figure}
\begin{center}
\includegraphics[width=1.\textwidth]{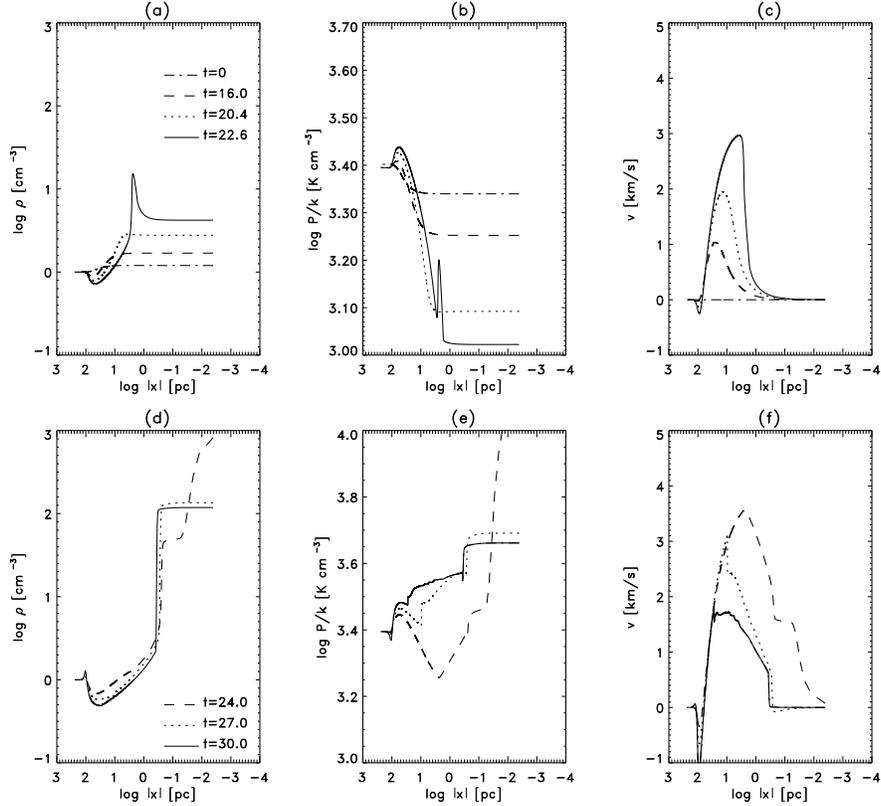}
\end{center}
\caption[]{Same as fig.\ \ref{fig:den3} but for run DEN75. The first
part of the evolution is shown in the upper frames, and the rest
in the lower frames. Note the formation of a shock shortly after
$t=22.6$ (frame c), which then propagates outwards from the cloud. At
the same time, the density overshoots to over $55 \rho_{\rm isob}$
(frame d). After the formation of the cloud, the density relaxes to a
value $\sim 2.5 \rho_{\rm isob}$, due to the ram pressure of the still
infalling gas.}
\label{fig:den75}
\end{figure}
Its evolution is shown in figs.\ \ref{fig:den75} (density,
pressure and velocity profiles) and \ref{fig:den75_Prho} (evolution on
the $P$-$\rho$ plane). In this case, the evolution is significantly
different. As dictated by the smaller growth rates of larger-scale
perturbations, run DEN75 requires 30 Myr to complete the formation of a
cloud, but moreover, throughout the first part of its development, the
condensation proceeds along the thermal equilibrium curve (see the first
four panels of fig.\ \ref{fig:den75_Prho}), developing locally supersonic
velocities in the process (maximum Mach number $\sim 1.2$) that cause a
strong overshoot. Thus, this condensation transiently reaches densities
$\sim 55 \rho_{\rm isob}$. At the time of maximum compression, a strong
shock is produced at the cloud boundary that propagates outwards from
it. This shock weakens quickly as it propagates into the low density
medium, but it has the important effect of heating the still-infalling
gas, bringing it out of thermal equilibrium and closer to isobaric
conditions (cf. last two
panels of fig.\ \ref{fig:den75_Prho}), and reversing the velocity gradient.
This shock is located at the peak of the velocity for the various times
shown in panel ($f$) of fig.\ \ref{fig:den75}, and is seen also in
the pressure (see panel ($e$) of the same
figure). By the end of the simulation, the accretion ram pressure 
is still high enough that the condensation has $\rho \sim 3 \rho_{\rm
isob}$, and this value decreases extremely slowly with time. Again, as 
in run DEN3, the infalling gas is mostly in the ``unstable'' density
range, in this case with a mass of almost twice that in the cloud. Within 
the infalling region, the pressure and density gradients have the same sign, so
this gas again has no further tendency to fragment. We have found from other 
simulations that the qualitative behavior of run DEN75 occurs down to
initial fluctuations with FWHM=15 pc.

 From the evolution of these two simulations, we conclude that
large-scale ($\eta \ll 1$) entropy perturbations have such a dynamic
evolution that their final central density and pressure are larger than
those corresponding to plain thermal-pressure equilibrium with the
diffuse phase, and moreover require such long times to
evolve (over 20 Myr to the occurrence of the large density overshoot),
that they are unlikely to complete their evolution before being
disrupted by other perturbations in the real ISM, such as passing shock
waves, or simply, general turbulent fluctuations. Small-scale entropy 
perturbations ($\eta \gtrsim 1$), on the other hand, adhere better to
the paradigm of forming near pressure-equilibrium condensations,
although we have seen that a significant fraction of the mass
still lies in the unstable range after the cloud has formed, and is accreting
onto the condensation, causing the presence of (weak) accretion fronts
(rather than contact discontinuities) at the cloud boundaries that only
subside asymptotically in time. Since the evacuation of the low
density regions must proceed in times of order of the sound crossing
time, the final fraction of mass in the ``unstable'' density range,
in the more realistic case of multiple fluctuations, should depend on
their number. In fact, we have performed simulations with a
full spectrum of initial fluctuations, and in those cases the final
unstable fraction may be much lower, although still times $\gtrsim 15$ 
(respectively, 8) Myr are required to evacuate the unstable range when the
minimum perturbation size is 12.5 (respectively, 1.25) pc.
More importantly, however, small-scale perturbations behave very
differently when they are quasi-adiabatic rather than quasi-isobaric, as
we discuss in the next section. 

\section{The case of velocity fluctuations} \label{sec:vel_fluc}

As mentioned in \S \ref{sec:adiab_pert}, velocity
fluctuations are likely to be the most representative of the actual
situation in the turbulent ISM, because in a continuum any density
fluctuation must originate from compressive or expansive motions. Such
motions are readily available in a compressibly turbulent medium. When
the cooling time is long ($\eta \gg 1$), these compressions/rarefactions
heat/cool the gas adiabatically, and the perturbations are then isentropic,
which, as discussed in \S \ref{sec:adiab_pert}, are stable in the
atomic ISM in the short-wavelength limit. As also mentioned in that
section, in the case of velocity fluctuations, the dynamical
time in $\eta\equiv\tc/\td$ is given by $\td=\min\{\ts,\tu\}$, where
$\tu$ is the turbulent crossing time, which is in general also a
scale-dependent quantity. 

Several studies \cite{Muf74,HP99,KI00} have investigated the possibility
of triggering TI through the nonlinear compression, either by strong shocks or
large-scale, large-amplitude compressive waves of the warm {\it stable} phase,
concluding in general that triggering TI off the stable phase is
possible for strong enough compressions, with the possibility of even
forming molecular hydrogen in the collapsed region \cite{KI00}. However, 
these studies have assumed that the gas has already previously segregated into
phases. It is our interest now to discuss the extent to which such
segregation can be achieved, starting from unstable conditions. Therefore,
in this section we describe the evolution of a thermally unstable
medium (with respect to the isobaric criterion) subject to
random velocity forcing, as originally presented in Paper I. 

We take uniform-density initial conditions, and
apply a random forcing $f$ of the form
\begin{equation} \label{eq:forcing}
f(x,t)= {\rm Re} \left[N \exp \left(ik\left(t\right)
x+i\phi\left(t\right)\right)\right], 
\end{equation}
where $k(t)$ is a time dependent wavenumber, $\phi(t)$ is the
phase and $x$ the position. Following \cite{Bra01}, we
take $N=f_{0}c_{\rm s}\left(k(t)c_{\rm s}/\delta t\right)^{1/2}$,
where $f_{0}$ is a constant factor and $\delta t$ is the length
of the timestep. The values of
$k/(2\pi/l)$, where $l$ is the box length, and of $\phi$ are selected at
each timestep randomly in the ranges $\left[3,10\right]$ and
$\left[0,2\pi\right]$, respectively. The positive exponent (1/2) in $N$
implies that strongest forcing occurs at the highest wavenumbers of the
forced range, so that the energy-injection scale scale
is $\lambda_{\rm i}=l/10$. We have chosen this forcing for two main
reasons. One, it mimics the small-scale stellar forcing acting in the
ISM, and, two, it allows us to maintain the desired rms Mach
numbers at the (small) scales of interest, since this is difficult to
achieve with pure large-scale forcing. We do not consider
decaying-turbulence situations, as we are ultimately interested in
models of the ISM, which is subject to continued energy-injection
processes.

We have performed nearly 20 simulations varying the box size and the
scale and amplitude of the forcing (see Paper I for details). From
them we conclude that, for 
the average conditions of the ISM, the presence of
turbulent motions with small enough sizes ($\sim 0.3$ pc) and moderate
amplitudes (${\mathcal M}_{\rm rms} \gtrsim 0.3$) such that $\eta$ is
maintained above unity, condensations do not appear. We understand this as a
consequence 
of the turbulent crossing time becoming shorter than the growth time
of the condensations, allowing the turbulent fluctuations to both
disrupt the incipient condensations and to more than compensate cooling through
the heating from shocks and adiabatic compression; i.e., the
perturbations become adiabatic, and therefore stable according to the
linear analysis in the limit $\eta \gg 1$.

In general, we conjecture 
that the presence of velocity fluctuations in the ISM, even if unable to  
completely suppress the development of condensations, may increase the
fraction of gas in the ``unstable'' temperature range. Note also
that for small-scale velocity fluctuations, the evolution is {\it not}
along the thermal-equilibrium curve, but rather intermediate between
adiabatic and isobaric. Thus, the density range determined by the
``unstable'' temperature range under these conditions does not exactly
coincide with the unstable density range under thermal equilibrium, as
given in \S \ref{sec:quali} and fig.\ \ref{fig:Peq}. An interesting
possible application for observationally measuring the actual
thermodynamic state of the atomic ISM is mentioned in \S
\ref{sec:conclusions}. 

\section{The magnetic pressure in turbulent media} \label{sec:magn_pres}

We now make a pause in the discussion of the thermal instability and
consider the character of the magnetic pressure in turbulent flows, in
order to assess the possibility that it may supplement the thermal pressure
in the ISM, and thus contribute to weaken the effects of TI. Note that
in this section we make no attempt to discuss TI in the presence of a magnetic
field. This has been discussed by a number of authors (e.g.,
\cite{Fie65,OMB82,Loe90,HP00}). Instead, here we investigate the nature
of magnetic pressure in fully turbulent compressible,
magnetized isothermal flows. Several works have considered
this regime as well, both numerically (see, e.g., the reviews by Mac Low,
Ostriker and Nordlund in this volume, and references therein), and
theoretically \cite{LG01}. In particular, the numerical simulations of 
references \cite{PN99} and \cite{OSG01} (see also \cite{PVP95} for the
nonisothermal case) reported a lack of correlation
between the density and the magnetic pressure, $B^2$, where $B$ is the
magnetic field strength, at low and intermediate densities in cases in which
the magnetic $\beta$ parameter\footnote{The magnetic $\beta$ should not
be confused with the exponent $\beta_{ij}$ of eq.\ 
\ref{eq:cooling}}, equal to the ratio of thermal to
magnetic pressure, is $\sim 1$. Moreover, recent observational data
(see, e.g., Crutcher, Heiles \& Troland, this volume) suggest a similar
lack of correlation at densities below $\sim 1000$ cm$^{-3}$ in the
ISM. Paper II has attempted to understand the origin of this
decorrelation in terms of the so-called ``simple'' MHD waves, and
discussed its implications on the role of $B^2$ as a 
pressure. In this section we briefly summarize the results of that
paper.

We consider isothermal MHD flows in ``1+2/3'' dimensions, or slab
geometry. The direction $x$ is referred to as the direction of wave 
propagation. In this setup, $b_x$, the field component along $x$, is
constant. We denote by $b$ the magnitude of the field component
perpendicular to 
$x$. The initial, uniform magnetic field is chosen to lie in the
$(x,z)$ plane, at an angle $\theta$ from the $x$ axis, so that $b_x=\cos 
\theta$ at all times. The treatment in this section is entirely in
non-dimensional units, so that the 
parameters characterizing the flow are the
sonic and Alfv\'enic Mach numbers of the velocity unit,
denoted $M_{\rm s}$ and $M_{\rm A}$, respectively, and the
propagation angle $\theta$. The plasma beta is then
$\beta=M_{\rm A}^2/M_{\rm s}^2$.

\begin{figure}[b]
\begin{center}
\includegraphics[width=.7\textwidth]{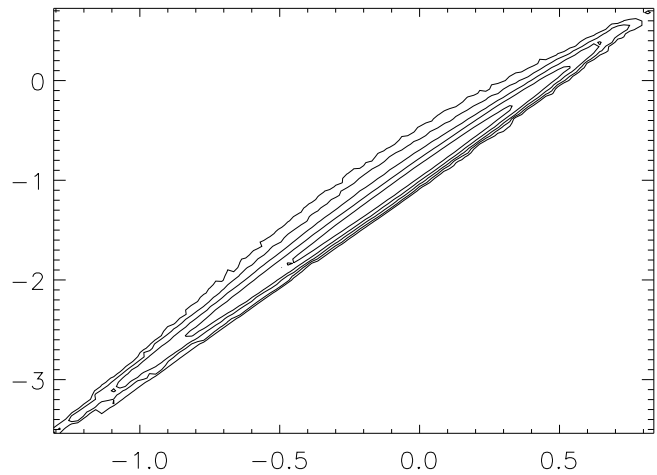}
\end{center}
\caption[]{Magnetic pressure-density correlation, indicated by the
two-dimensional histogram of points in log-log coordinates, for a
simulation with  
a magnetic field perpendicular to the direction of propagation (i.e.,
$\cos \theta = 0$), and forcing parallel to this direction. This
configuration allows only the existence of the fast mode of nonlinear
MHD waves. The run has an rms field fluctuation $\delta B/B = 0.62$ and 
$\delta \rho/\rho = 0.62$. The
rms Alfv\'enic Mach number $\tilde M_{\rm A}=5.2$.
 The magnetic pressure is seen to scale as $\rho^2$.} 
\label{fig:cos0Ma2}
\end{figure}

``Simple'' MHD waves (see, e.g., \cite{LL87,Man95}) are finite-amplitude
solutions of the equations, characterized by
the property that all variables can be expressed as wave ``profiles'',
i.e., as a function of a single one of them (say, the density) as in
the case of linear MHD waves. The same well-known modes of the linear
case, i.e., Alfv\'en, fast, and slow, exist in the case of simple
waves. Only the latter two are associated with the density fluctuations.
The propagation velocities of the modes are given by \cite{LL87,Man95} 
\begin{equation}
v_{\pm}^2={1\over 2M_a^2 \rho}\left ({B^2}+{\beta \rho
} \right ) \left ( 1\pm \sqrt { 1-{4\beta b_x^2\rho \over
(B^2+\beta \rho)^2}} \right ) \label{eq:fs-speeds}
\end{equation}
and
\begin{equation}
v_A=\pm {b_x\over M_a \rho^{1/2}}, \label{eq:alf-speed}
\end{equation}
where $v_\pm$ denotes the speed of the fast ($+$) and slow ($-$) modes,
and $v_A$, that of the Alfv\'en mode. $B^2=b_x^2+b^2$ is the total field
strength. 

After manipulating the equations to obtain the wave profiles, one 
finds, in particular, for the dependence of the field with density,
\begin{equation}
{d\over d\rho}{b^2\over 2}={d\over d\rho}{B^2\over
2}=(M_a^2 v^2-\beta).\label{eq:profileb} 
\end{equation}
In the limit when $4\beta b_x^2\rho \ll (B^2+\beta\rho)^2$, this
equation can be simplified and integrated using (\ref{eq:fs-speeds}) to
give the dependences for the fast and slow modes as
\begin{eqnarray}
{b^2} \approx C - 2 {\beta \rho} ~~~~\hbox{(slow
mode)} \label{eq:b_rho_slow}\\
{B^2} \propto {\rho^2}~~~~~~~\hbox{(fast mode),} \label{eq:b_rho_fast}
\end{eqnarray}
where $C$ is a constant.
These equations essentially give the behavior of magnetic pressure with
density for the two modes in the limit mentioned above. This condition
is generally satisfied, except when $\beta\rho \approx b_x^2$ and
simultaneously $b^2 \ll b_x^2$, i.e., when $b_x$ is not too small, for
$\beta\rho$ of order unity and small field distortions.

Several points are noteworthy about (\ref{eq:b_rho_slow}) and
(\ref{eq:b_rho_fast}): 1) The $\beta$ weighting in the second term
of the RHS of (\ref{eq:b_rho_slow}) implies that at low $\beta$,
the slow mode produces large density fluctuations even when the field 
fluctuations are small. 2) The total pressure,
$P_{tot}=\frac{b^2}{2M_A^2}+\frac{\rho}{M_s^2}$,  
is rougly constant in the slow mode.  3) Most importantly, the
pressures from the two modes depend very differently on density. One
can then expect that, in the large fluctuation amplitude case (i.e., the 
fully nonlinear regime), {\it the
particular value of the magnetic pressure of a fluid parcel
will not be uniquely determined by its density, but instead, that it will
depend on the detailed history of how the density fluctuation was
arrived at}, causing a lack of correlation between the magnetic pressure 
and the density. 

The latter suggestions have been tested in Paper II by means 
of numerical simulations with random forcing (actually, an acceleration)
applied on wavenumbers 1-19 to all three velocity components or to
only the perpendicular ones.  
Choosing the direction of the forcing and of the uniform magnetic
field allows us to highlight either one of the slow and fast
modes. Figure \ref{fig:cos0Ma2} shows the $b^2$-$\rho$ correlation by
means of isocontours of the two-dimensional histogram in 
the $\log (b^2/ 2 M_A^2)$--$\log (\rho)$ plane of the points in a
simulation with 4096 grid 
points, forcing applied on all three velocity components, and the initial
magnetic field perpendicular to the direction of propagation. This is a
case in which only the fast mode exists, consisting of a pure
magnetosonic wave, and the correlation exhibits the well known $\rho^2$
behavior of magnetic pressure in this case. This result holds independently
of the value of $M_A$.

\begin{figure}[b]
\begin{center}
\includegraphics[width=.7\textwidth]{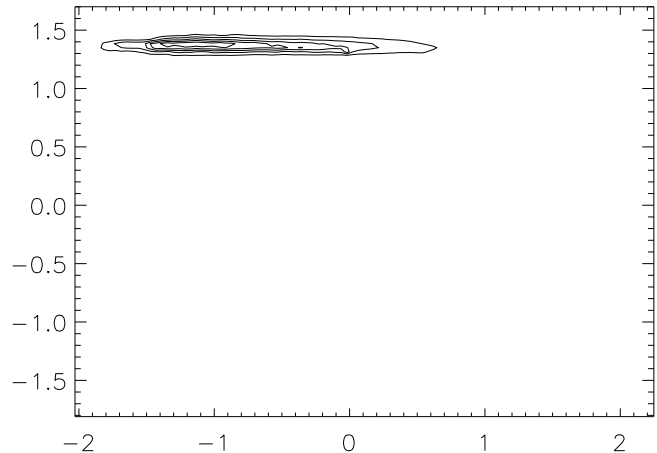}
\end{center}
\caption[]{Magnetic pressure-density correlation for a simulation with 
$\cos \theta = 0.1$, and forcing perpendicular to the propagation
direction, with $\delta B/B = 0.32$ and $\tilde M_A= 0.48$. This
configuration highlights the slow mode. The magnetic 
pressure is seen to remain virtually constant.}
\label{fig:cos01Ma015}
\end{figure}

\begin{figure}[b]
\begin{center}
\includegraphics[width=.7\textwidth]{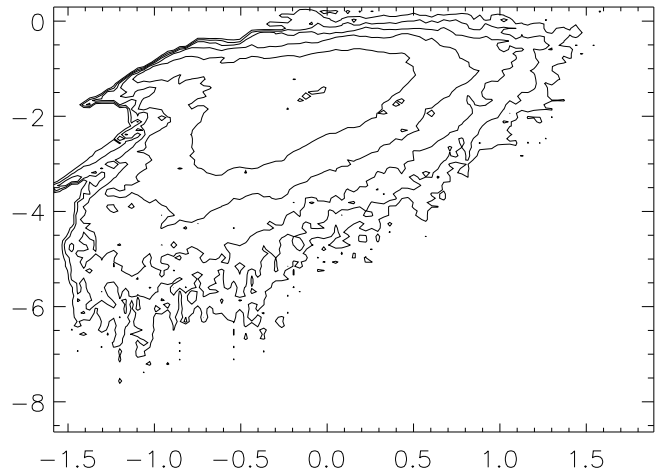}
\end{center}
\caption[]{Magnetic pressure-density correlation for a simulation with 
$\cos \theta = 0.1$, and forcing in the three
directions. This configuration also highlights the slow mode, but increases
the field distortions ($\delta B/B = 2.17$ and $\tilde M_{\rm A }= 7.29$). The
magnetic pressure is seen to decorrelate from the density, although it
appears to be bounded above and below by the slow and fast mode
dependences, respectively.}
\label{fig:cos01Ma6}
\end{figure}

Figure \ref{fig:cos01Ma015}, on the other hand, shows the correlation
for a run at the same resolution with the 
forcing perpendicular to the propagation direction, and the
magnetic field almost parallel to the forcing ($b_x=\cos \theta =
0.1$), at low Alfv\'enic Mach number ($M_{\rm A}=0.15$). In this case,
the density fluctuation production is dominated by the slow mode because 
$\tilde \beta\sim 0.007$
(c.f. (\ref{eq:b_rho_slow})). The near constancy of the total 
pressure in this case is evident in this figure. 
In this simulation, one observes large oscillating density clumps which do
not merge nor pass each other, mostly anti-correlated with $b^2$. 
However, for this same
field configuration, as $M_{\rm A}$ is increased, the field
becomes more strongly distorted, and the fast mode starts acting on the
perturbed field. The fluctuations become more random and superpose
each other, with the presence of fast shocks and a correlation between 
$\rho$ and $b^2$. The result is that at high $M_{\rm A}$ both modes are
actively producing density fluctuations, and the correlation between
magnetic pressure and the density is lost, as seen in fig.\
\ref{fig:cos01Ma6}. The situation can be 
idealized by assuming that the density of each fluid parcel is arrived
at through a random sequence of slow and fast waves, which is different
for each parcel. 

The case of parallel propagation is also interesting to mention as it
illustrates the complexity of this problem. For example, 
the picture of non-interacting clumps observed at small
$M_A$ (i.e. in presence of weak field distortions) 
and  for large angles is also observed for parallel propagation
in the case of large $M_A$. But in that case the clumps form inside strong
slow shocks. The magnetic field intensity is the weakest inside the
clumps which cannot merge due to the high external magnetic pressure.
In the general 3D case, we expect all angles between the magnetic field
and the propagation direction to be present, and therefore a
representative case would be one in which the field is at $45^\circ$
from the propagation direction. In this case, we recover the trend of a
roughly constant magnetic pressure at small $M_{\rm A}$ and an increased
scatter between the magnetic pressure and density as $M_{\rm A}$ is
increased (not shown). 
It is found that the level of density fluctuations greatly depends on the 
dominant mode, has no specific relation with $\beta$ and is actually the 
largest when $B$ is strong (and thus slightly perturbed). Density PDF is close
to a log-normal when $B^2$ and $\rho$ are not correlated, and shows an excess
at small density when slow waves dominate.

We conclude from this section that the scatter between magnetic pressure 
and density found in simulations and in observational data can be
understood in terms of the different dependence of the magnetic intensity
on density for the slow and fast modes of simple nonlinear MHD waves,
and of the random sequence of these that a fluid parcel experiences as it
evolves in a fully turbulent regime. Moreover, a number of important
implications emerge from this lack of correlation. First, it suggests
that modeling magnetic pressure by means 
of a polytropic dependence on density may not be adequate in the fully
turbulent case. This relation is seen to apply 
when scale separation is preserved
between small-scale Alfvén waves and large-scale density perturbations.
Second, the magnetic ``pressure'' does not really act as 
a pressure, as it does not behave as a restoring force in
general. Instead, it acts more as a random forcing. Third, the latter
point suggests that magnetic pressure should not be effective as a
substitute for thermal pressure when the latter behaves ``in reverse'' in 
thermally unstable situations.
A final note is that the isothermality of the flows considered here
is of no relevance to the results, which are thus expected to apply
equally to non-isothermal flows.

\section{TI in models of the ISM} \label{sec:full_ISM}


In the previous sections we have discussed the nonlinear development of
fluctuations of various kinds and sizes in the presence of TI under the
isobaric criterion, and the nature of magnetic pressure in turbulent
media. With this background we can now proceed to discuss the
behavior of numerical models of the ISM incorporating the magnetic
field, self-gravity, rotation, shear, and stellar-like (localized) energy
injection, in the presence of isobaric TI.

As mentioned in the introduction, the classic two- and three-phase
models of the ISM were based on the principles of thermal and pressure
equilibrium, and thus did not predict the existence of significant
amounts of gas in-between the phases. Instead, the time-dependent model
\cite{GKM74} did. Furthermore, observations do not clearly support a
sharp phase segregation either; instead, they often have found evidence
of significant 
amounts of ``lukewarm'' gas at temperatures intermediate between those
of the cold and warm phases \cite{DST77,KSG85,SF95,FS97,Hei01}.
Thus, it is important to perform numerical simulations that quantify
this fraction and that allow us to determine whether sharp phase
segregation is expected in the ISM, or whether it is more likely a
continuum.

Numerical simulations of the ISM including radiative heating and cooling 
as well as stellar-like energy injection have been 
performed by a number of authors, starting from the pioneering work of
Bania \& Lyon \cite{BL80}, and continuing with references
\cite{CP85,CB88,RB95,VPP95,PVP95,Elm97,GI97,GP99,Kor99,dAv00,VGS00,GVSS01,KBM01,MBKA01}. These works have included different
amounts of physics in the simulations, such as the magnetic field,
self-gravity, galactic disk rotation, etc. However,
the role of TI had been discussed only in passing until recently. Bania
\& Lyon, using two-dimensional (2D) simulations of a 180-pc square
region on the Galactic plane, at a resolution of $40\times 40$
pixels, including randomly-positioned stellar-like energy sources,
pointed out that the inclusion or not of a thermally unstable range had
little effect on the resulting structure. Elmegreen
\cite{Elm97} pointed out that high-resolution 1D MHD simulations forced
with nonlinear magnetic waves can form hierarchical cloud/intercloud
structure both if the cooling functions used have a single 
or double stable equilibria. The 2D non-magnetic simulations of
Gerritsen \& Icke \cite{GI97}, and the 3D MHD ones of Korpi et
al. \cite{Kor99}, although not specifically
aimed at this issue, already pointed towards the existence of
unstable gas. \VS, Gazol \& Scalo \cite{VGS00} (hereafter Paper III)
first discussed the interaction of TI with the turbulent motions in the
ISM, suggesting that at large scales (simulation size of 1 kpc)
the signature of TI in the mass density histogram is erased when
small-scale forcing mimicking expanding HII regions placed at the
density peaks is used in low-resolution ($128\times 128$) 2D simulations
including the magnetic field, self-gravity and rotation. 
Gazol et al.\ \cite{GVSS01} (hereafter Paper IV) then reported that
the temperature histogram in similar simulations at higher resolution
($512 \times 512$) contains roughly half the mass at
unstable temperatures. Recently, Mac Low et al.\ \cite{MBKA01} (see also 
Mac Low, this volume) have presented 3D MHD simulations spanning the whole
range of temperatures existing in the ISM, including the hot ($T \gtrsim 
10^6$ K) gas, showing that the effect of supernovae is to introduce
large fluctuations (by 2--3 orders of magnitude) in the thermal
pressure of the ISM, in contrast with the multiphase models of the ISM
\cite{FGH69,MO}, and more in agreement with the time-dependent ones
\cite{GKM74}, and with the recent observational study of Jenkins \& Tripp
\cite{JT01}. Finally, in an analytical treatment of the effective
equation of state of the ISM and the power spectrum of the energy
sources, Norman \& Ferrara \cite{NF96} concluded that the traditional
multiphase description should be replaced by a ``continuum of phases''.

A related line of study has been that taken by Koyama \& Inutsuka
\cite{KI01} and Kritsuk \& Norman \cite{KN02}. Both of these groups have 
recently considered the generation of turbulence in flows in more than
one dimension, as a consequence of the nonlinear development of
TI. Koyama \& Inutsuka considered a shock-compressed layer in 2D between a
fast flow with diffuse-gas properties ($n=0.6$ cm$^{-3}$, $T=6000$ K)
and a hot gas region, showing that the layer fragments into small
cloudlets that have supersonic velocity dispersions with respect to the
warm medium in which they are embedded, and coalesce to form larger
units. Kritsuk \& Norman considered the 3D development of the instability
alone, very far from thermal equilibrium ($n=1$ cm$^{-3}$, $T=2\times
10^6$ K), so that the gas is initially unstable under the isochoric
mode and the cooling times are very short ($\sim 0.3$ Myr). They again
found that the development of TI generates turbulence, in which roughly
15\% of the mass is in the unstable regime. They point out, however,
that this turbulence is decaying, because it has no other energy sources 
than the development of TI itself. 

However, none of the works reporting a fraction of unstable gas have
discussed in detail the possible
suppression of the instability by numerical limitations, raising
a concern that perhaps the presence of unstable gas, and therefore the
lack of sharp phase transitions in the simulations, are numerical
artifacts. In the remainder of this section we first present new results
concerning the numerical issues in detail, in order to assess the
validity of the ISM simulations and interpret their results. We then
present new simulations that, based on 
the numerical considerations, provide reasonable evidence that
significant amounts of gas at unstable temperatures should be expected
in the atomic ISM.

The simulations presented in this section have been performed using a
pseudo-spectral scheme, described in detail in \cite{VPP96}, to solve
the MHD equations in the presence of heating, cooling, stellar-like
energy sources, and self-gravity, using a hyperviscosity scheme and
including a mass diffusion term with coefficient $\mu$ (cf.\ eq.\
\ref{eq:1Dcont}) and a thermal diffusion term with a constant
coefficient $K=K_0$ (cf.\ eq.\ \ref{eq:int_en}). In non-dimensional
form, the equations read
\begin{equation}
{\partial\rho\over\partial t} + {\nabla}\cdot (\rho\u) = \mu
\nabla^2 \rho, \label{eq:pse_cont}
\end{equation}
\begin{eqnarray}
{\partial\u\over\partial t} + \u\cdot\nabla\u =
-{\nabla P\over \rho}
- \Bigl({J \over M}\Bigr)^2 \nabla \phi + {1 \over \rho}
\bigl(\nabla \times {\bf B}\bigr) \times {\bf B} \nonumber \\
~~~~~~~~~~~~~~~~~~~~ - 2 \Omega \times \u - {\nu_8} {\nabla^8\u}+
\nu_2 (\nabla^2 \u + \frac{1}{3} \nabla \nabla \cdot \u),\label{eq:pse_mom}
\end{eqnarray}
\begin{eqnarray}
{\partial e\over\partial t} + \u\cdot\nabla e = -(\gamma -1)
e\nabla\cdot \u + K {\nabla^2e\over \rho} +
\Gamma_{\rm d} + \Gamma_{\rm s} - \rho \Lambda, \label{eq:pse_ener}
\end{eqnarray}
\begin{equation}
{\partial {\bf B}\over\partial t} = \nabla \times (\u \times {\bf B})
- {\nu_8} {\nabla^8{\bf B}} + \mu \nabla^2 {\bf B}, \label{eq:pse_magn}
\end{equation}
\begin{equation}
\nabla^2 \phi=\rho -1, \label{eq:pse_poisson}
\end{equation}
\begin{equation}
P=(\gamma-1)\rho e, \label{eq:pse_eqstat}
\end{equation}
\begin{equation}
\Gamma_{\rm d}({\bf x},t)=\Gamma_{0},
\label{eq:pse_gammad} 
\end{equation}
\begin{equation}
\Gamma_{\rm s}({\bf x},t) =
\cases{{\rm \Gamma_1}   & if $\rho({\bf x},t_0) >
        \rho_{\rm cr}$ \cr
        & and $0 < t-t_0 < \Delta t_s$\cr
        0               & otherwise,\cr}, \label{eq:pse_gammas}
\end{equation}
where $M$ is the Mach number of the velocity unit, taken equal
to unity, $J=l/L_{\rm J}$ is the box size in units of the Jeans length,
$\phi$ is the gravitational potential,
$\Omega$ is the Galactic disk rotation rate, $\nu_2$ and $\nu_8$ are
respectively the second-order and eight-order (hyperviscosity)
coefficients, $e$ is the specific internal energy, $K$ is the thermal
conductivity, $\Gamma_{\rm d}$ and $\Gamma_{\rm s}$ are respectively the
diffuse background heating rate and the local stellar heating rate, and
$\Gamma_0$ and $\Gamma_1$ are constants. The equations are
non-dimensionalized to the box size $L_0$, the velocity unit $u_0=c$, the
temperature unit $T_0=10^4$K and the magnetic field unit $B_0=5 \mu$G. 

The cooling rate $\Lambda$ is still given by eq.\ (\ref{eq:cooling}),
although the nondimensional coefficients are now given in Table
\ref{tab:pse_cool_coef}, together with the non-dimensional heating rates,
for three different box sizes, at the same temperature unit.

\begin{table}
\caption{Cooling function parameters for ISM simulations}
\begin{center}
\renewcommand{\arraystretch}{1.4}
\setlength\tabcolsep{5pt}
\begin{tabular}{cccccccc}
\hline\noalign{\smallskip}
Box size (pc)	& $C_{12}$& $C_{23}$& $C_{34}$& $C_{45}$ & $\Gamma_0$ & $\Gamma_1$ \\
\hline\noalign{\smallskip}
10		& 56.2	& 0.462	& 0.102	& 0.474	& $4.57 \times 10^{-2}$ & 20\\
150		& 845.	& 6.95	& 1.54	& 7.14	& 0.688	& 38.7		\\
1000		& $5.62 \times 10^3$& 46.2& 10.2& 47.4	& 4.57	& 250	\\
\hline
\end{tabular}
\end{center}
\label{tab:pse_cool_coef}
\end{table}

The ISM simulations are started with uncorrelated gaussian random
fluctuations in all variables and amplitudes of order unity, with
characteristic scale $\sim 1/8$ of the box size. The initial magnetic
field has a uniform component of 1.5 $\mu$G on the $x$-direction, and an 
rms fluctuation amplitude of 4.5 $\mu$G. The energy injection mechanism
consists of small-scale ($\sim 10$ pixels across) heat sources turned on
at sites where the density exceeds a certain threshold (chosen within the cold
stable branch of the density range), that remain on for $\Delta t = 6$
Myr, except in the simulations with box size = 10 pc, for which the time 
unit is too short, and a star would remain on for more than half the
duration of the simulation. In this case we have shortened $\Delta t$ by 
a factor of 10, and increased the energy injection rate by roughly the
same factor. These sources are intended to mimic
the effect of ionization heating from OB stars in HII
regions. Supernova-like sources are not included because of limitations
of the numerical scheme to handle very strong shocks.

\subsection{Numerical considerations} \label{sec:numerical}

The numerical simulation of thermally unstable turbulent flows presents
a significant numerical challenge because it requires solving
simultaneously the regions of sharp gradients occurring in the
immediate neighborhood of clouds and the quasi-linear development of
perturbations in the more distant, relatively quiescent, unstable medium 
mediating the clouds and the warm, stable, diffuse phase, the question
being whether this medium fragments, to finally end with a state in
which the unstable gas virtually disappears from the simulation. The
simultaneous solution of both regimes is important, because stellar
energy injection recycles gas from the dense phase into the warm phase.

The simulation of regions with strong shocks requires the use of artificial
viscosities and diffusivities in order to spread out (``capture'')
shocks over a few grid points. Unfortunately, such diffusivities also
have the effect of damping the growth of perturbations in the relatively 
smooth regions, because they artificially increase the Field length (\S
\ref{sec:entropy_pert}), reducing the range of unstable scales, as
well as their growth rates, in the unstable gas. This problem is also
present when finite-difference schemes, which produce numerical
diffusivity, are used. This was not a problem, however, in sections
\ref{sec:pure_TI} and \ref{sec:vel_fluc} because of the very large
resolutions used there, at the expense of using a 1D approach.

The Field length $\lf$ has the unfriendly (for our purposes)
property of being much larger than the diffusive scale (understood as
the molecular mean free path) when the latter is very small, since it
can be easily shown \cite{Fie65} that the wavenumber associated to the
Field length satisfies
\begin{equation}
k_{\rm F} = \left[\frac{\mu~(\gamma-1)}{R c} k_K |N_P| \right]^{1/2},
\label{eq:Field_wavenumber}
\end{equation}
where $k_K$, given by eq.\ (\ref{eq:k_K}), is the wavenumber associated
with the molecular mean free path. Thus, $k_{\rm F}$ only grows as the square
root of $k_K$. In a numerical simulation,
the scale associated with the artificial thermal diffusivity $\lambda_K
= 2 \pi/k_K$ (cf.\ \S \ref{sec:entropy_pert}) plays the
role of the molecular mean free path, and therefore 
the growth of perturbations with wavelengths  $\lambda_K <
\lf$ can be artificially suppressed if the numerical $\lf$ is much
larger than the real $\lf$ in the atomic ISM. In other words, much of
the resolution in a simulation is ``wasted'', in the sense that
intermediate-wavelength perturbations will be damped even if the
diffusive scale is comparable to the smallest resolved scale, as we will 
show below.

On the other hand, the pseudo-spectral numerical scheme we use in
simulations of the ISM has the advantage that
it produces no numerical diffusivities at all (the spatial derivatives are 
calculated exactly in Fourier space, rather than through
finite-difference approximations). Thus, all artificial diffusivities
are included explicitely in the equations, and can be controlled through 
their associated coefficients. This allows us to perform simulations of
the non-diffusive case in the linear and weakly nonlinear regimes
(recall the diffusivities are only needed to smooth out strong
gradients) and of diffusive cases, so that we can both test the code
against the predictions of the linear analysis, and then measure the
effect of the diffusivities precisely. We have found that actually the
mass diffusion has a much stronger damping effect on the growth of
perturbations than heat conduction and the viscosity. The value of the
thermal diffusivity used in the ISM 
simulations discussed below, $K=6.6 \times 10^{-3}$, was found to not
affect the growth rate by more than 10\% for the wavenumbers tested
below. A similar situation holds for the viscous and hyperviscous
coefficients.

We have therefore performed many simulations at low resolution
($128\times 128$) to measure the growth rates of
pure sinusoidal isobaric perturbations of various wavelengths (from 1 to 
1/32$\times$ the box size) as a function of the mass diffusion
coefficient, in order to investigate at what point the perturbation
growth is suppressed and thus full ISM simulations using those values
cannot be trusted anymore concerning the instability of regions of sizes 
comparable to those perturbations. Note that higher resolution is not
necessary for this purpose, as all that is needed is to resolve
the perturbations themselves and their initial growth. Higher
resolution is only necessary in the fully nonlinear case, to resolve
shocks while still having a large range of scales between the simulation
size and the scale of shock-spreading.

Together with the theoretical growth rate, figure \ref{fig:wvsk} shows
the growth rates (defined as the inverse of  
the $e$-folding time for linear -- 2.5\% amplitude -- perturbations)
measured in simulations of a region of 150 pc on a side as a function of
wavenumber and of the mass diffusion coefficient. These rates can be
compared with the solid line, which gives the solution of the dispersion 
relation (\ref{eq:disp_rel}) as a function of wavenumber. A good
agreement is seen between the theoretical and numerical rates for
the zero-diffusivity case. For non-zero diffusivity, the numerical rates
are seen to decrease significantly, by roughly a factor of 1.5 at $\mu=4
\times 10^{-3}$ and $\lambda=1/8$ of the box size. For the same diffusivity,
perturbations of size 1/16 of the box size have a zero growth rate (the
perturbations remain static, without growing or dispersing). 

This damping effect is alleviated somewhat by noting that
larger-amplitude perturbations have larger growth
rates \cite{SMS72}. We have thus also computed the numerical growth rates for
perturbations of initial amplitude of 25\% in simulations of box
size $= 150$ pc (not shown), verifying the occurrence of larger growth
rates in this case. This is advantageous because then weakly nonlinear
perturbations of scales down to sizes $\sim 15$ pc should grow in
simulations with box sizes 150 pc at rates comparable to those of linear 
perturbations without any diffusivity. Nevertheless, we have found that zero
growth still occurs at the same scale (1/16 of the box) as for the linear
perturbations at roughly the same value of $\mu$. Any scales below
this are stabilized by the mass diffusion, and thus may be wrongly
interpreted as stable in a 150-pc ISM simulation using this value of the 
diffusivity.

\begin{figure}
\begin{center}
\includegraphics[width=.7\textwidth]{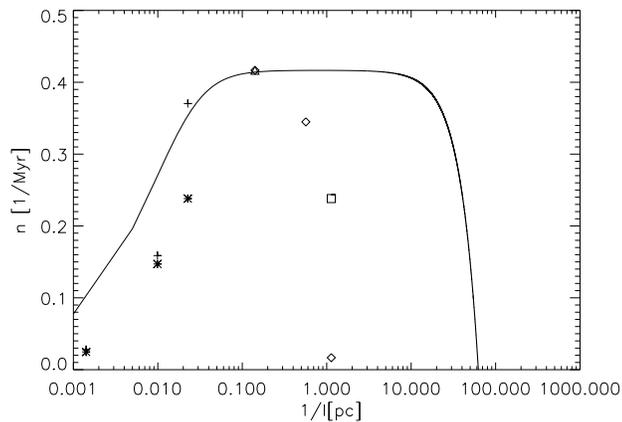}
\end{center}
\caption[]{
Numerical growth rates for simulations of the linear growth of 2.5\%
isobaric density perturbations
in boxes of sizes 10 and 1000 pc and various values of the mass
diffusion coefficient, superposed on the growth rate curve. {\it Crosses:} 
1000 pc, $\mu = 0$. {\it Asterisks:} Box size=1000 pc, $\mu = 0.008$. {\it
Triangles:} Box size=10 pc, $\mu = 0.0$. {\it Diamonds:} Box size=10 pc,
$\mu = 0.001$. {\it Squares:} Box size=10 pc, $\mu = 0.0005$. 
}
\label{fig:wvsk_10_1000}
\end{figure}

Figure \ref{fig:wvsk_10_1000} shows the numerical growth rates for 2.5\%
perturbations on simulations of box sizes 10 pc and 1 kpc on top of the
theoretical growth rate curve. Remarkably, it is seen that the mass
diffusion is much more effective in damping the growth of perturbations
for a small box size than for a large one. Specifically, a diffusivity
of $\mu = 0.001$ is enough to completely damp perturbations of size 1/8 of
the box when the box size is 10 pc, while a value $\mu=0.008$ still
allows growth of perturbations of size 1/16 of the box for a box size of 
1 kpc. This is actually easy to
understand because a larger box size is represented, in non-dimensional
units, by larger thermal coefficients (see Table
\ref{tab:pse_cool_coef}), while the diffusive coefficients are
independent of the physical box size, if the diffusive scale
is kept at a given fraction of the numerical box (in pixels). This
is similar to the effect on the Field length of the thermal
conductivity (eq.\
[\ref{eq:Field_wavenumber}]). Somewhat surprisingly, we conclude that, at
a fixed value of the 
mass diffusion coefficient, {\it a simulation with a larger physical box size
is more accurate than one with a small physical size.}

A final comment is that, from the results of this section, it appears
that the numerical scheme best suited for solving the problem at hand
(i.e., capturing strong gradients occurring in dense clouds while not
disturbing the growth of perturbations in the mildly turbulent diffuse
medium surrounding the dense clouds), is a pseudo-spectral scheme with a
position-dependent value of the diffusivities, as in eq.\
(\ref{eq:shock_visc}), so that strictly zero-diffusivity can be used in mild
regions, while still smoothing out the strong gradients in the
clouds can be achieved. We will 
attempt such an approach in forthcoming papers. However, our presently
available tools for modeling the full ISM problem only include
constant-coefficient diffusivities, and we will discuss results using
them in the next subsection. Note that not even
adaptive-mesh refinement schemes may be better suited for this problem,
because the small-amplitude perturbations need to be followed in the
diffuse medium in order to determine whether they grow spontaneously,
and this would require high refinement levels in relatively large volumes.

\subsection{Results} \label{sec:ISM_sim_res}

In the light of the results of \S \ref{sec:numerical}, we can now
proceed to present some numerical simulations of the ISM and assess
their reliability regarding the mass fraction in the unstable range.

In Paper IV, we presented a 2D simulation of the ISM with a 1-kpc 
box at a resolution of $512^2$ grid points with a mass diffusion
coefficient $\mu = 0.0075$. From fig.\ \ref{fig:wvsk_10_1000} it can be
seen that perturbations of sizes down to 1/16 of the box can grow at
this value of $\mu$ with a physical box size of 1 kpc, while
perturbations of size 1/32 of the box are damped. Thus, although this
box size is the one that provides the largest range of unstable scales,
it nevertheless does not reach down to the
fastest-growing scales. Thus, we have performed two more simulations, to 
cover the entire range of scales of interest: the first one with a box
size of 150 pc, and a very-high resolution of $1536^2$ grid points, in
order to be able to use a mass diffusion coefficient $\mu=0.003$ which,
according to the data in fig.\ \ref{fig:wvsk}, allows growth again of
perturbations of size 1/16 of the box, thus barely reaching the scales
of fastest growth ($\lesssim 10$ pc). The second simulation uses a box
of 10 pc, and $\mu = 5 \times 10^{-4}$ at a resolution of $512^2$, thus
allowing the growth of perturbations down to scales 1/8 of the box size.
We have opted for performing 2D simulations in order to maintain
relatively high resolutions, while still being able to capture the
complex vector interactions of the system.

\begin{figure}
\begin{center}
\includegraphics[width=.7\textwidth]{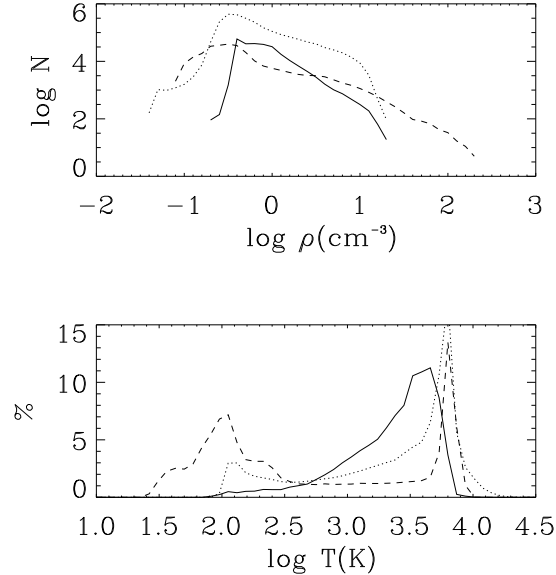}
\end{center}
\caption[]{{\it Top:} Density histograms for the ISM simulations
with box sizes 10 pc ({\it solid line}), 150 pc ({\it dotted line}) and
1000 pc ({\it dashed line}). A clear trend toward higher phase
segregation (more strongly bimodal shape) is seen at {\it large}
physical box sizes because the wave mode is unstable at large scales only.
{\it Bottom:} Temperature histograms, with the same labeling.
%
} 
\label{fig:pdfs}
\end{figure}

In fig.\ \ref{fig:pdfs} we show the density and temperature histograms
of these three runs after a stationary regime has been attained. It can
be seen that the density histograms do show signatures of TI, such as
slope changes and slight peaks at the stable densities, but
nevertheless, a sizeable fraction of the gas is in the unstable
regime. A similar result was reported by Kritsuk \& Norman \cite{KN02}.

Concerning the temperature histograms, we see that, in fact, as the
physical box size decreases, the temperature 
histogram has less pronounced spikes at the temperatures corresponding
to the stable phases, suggesting that phase segregation
is less pronounced as well. In the case of the 10-pc run, this can be an 
artifact of the reduced unstable range due to the mass diffusion.
However, this is not so for the 1-kpc and 150-pc runs, as both have
roughly the same range of unstable scales, and in fact with larger
growth rates in the case of the latter, because the corresponding
physical scales are smaller. Thus, these two runs suggest that in fact
the result is real, due to the decreasing value of $\eta$ as smaller
scales are considered, because in this case the perturbations are
adiabatic-like and are stable to first order. 
Taking the simulations at face value, the mass fractions in the
``unstable'' temperature range are 27\% for the 1-kpc run, 58\% for the
150-pc run, and 92\% for the 10 pc run. Again, we see the trend of a
{\it larger} fraction of unstable gas as smaller scales are considered.

\begin{figure}
\begin{center}
\includegraphics[width=0.7\textwidth]{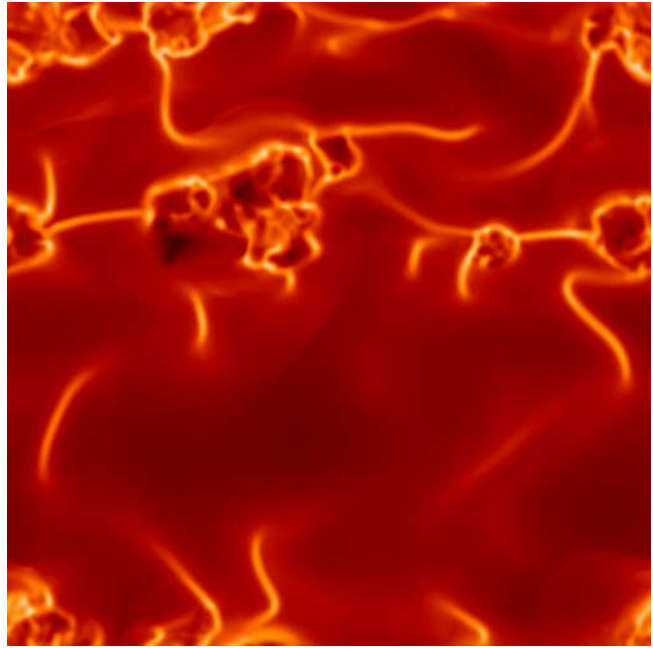}
\end{center}
\caption[]{Density field of the 150-pc simulation of the ISM at the same 
time for which the density and temperature histograms are
shown in fig.\ \ref{fig:pdfs}. The resolution is $1536^2$ grid points.}
\label{fig:150pc_den}
\end{figure}

\begin{figure}
\begin{center}
\includegraphics[width=0.7\textwidth]{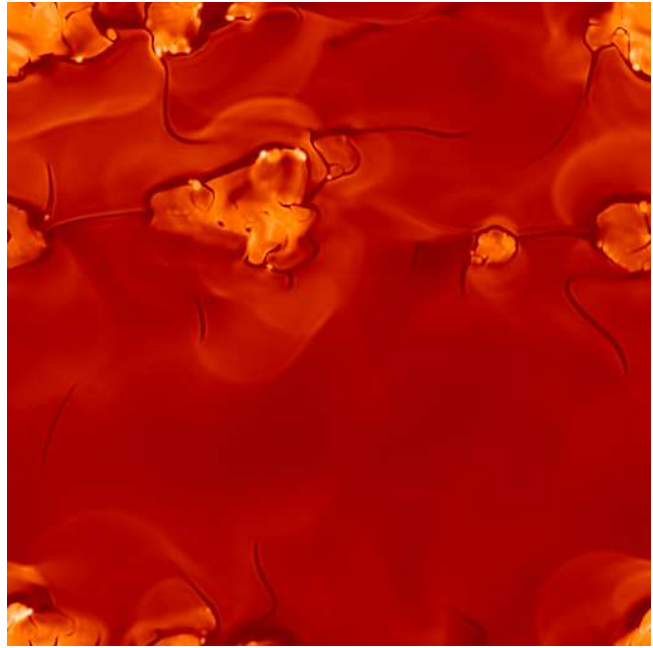}
\end{center}
\caption[]{Thermal pressure field of the 150-pc ISM simulation.
The highest pressures correspond to regions of ``star formation''
clustering, in which the heating from many stars combines
additively. Note the high-pressure regions around the clouds, left by
the weak shocks propagating into the intercloud medium.
}
\label{fig:150pc_pres}
\end{figure}

On the other hand, the effect of strong compressions continues to
promote the instability. Figures \ref{fig:150pc_den} and
\ref{fig:150pc_pres} respectively show the density
and pressure fields for the 150-pc run at a typical time. There it can
be appreciated that the density maxima (and the filaments connecting
them) correspond to pressure {\it minima}, indicating that these regions
are unstable, even when they are much smaller than the smallest linearly 
unstable scale allowed by the mass diffusion. This is because the
strong turbulent compressions locally increase the density and the cooling
rate, decreasing $\eta$. Thus, {\it these regions belong to the long-wavelength
(small-$\eta$) regime in spite of having small physical sizes}, and the
pressure behaves closer to the thermal-equilibrium curve. 
These small-scale, small-$\eta$ regions are analogous to 
the compression-induced instability of Hennebelle \& P\'erault
\cite{HP99}, except that in this case they have been pushed from the
large- to the small-$\eta$ regime by the compression, rather than from
the stable to the unstable regime. They also correspond to the
transition from isobaric to isochoric cooling as the density increases
described by Burkert \& Lin \cite{BL00} and Kritsuk \& Norman \cite{KN02}. 

It is worth comparing here the present results to those of Hennebelle \& 
P\'erault in somewhat greater detail, as those authors did not find
significant amounts of unstable gas in their simulations after the
condensation process ended. This is because their 
setup did not consider a globally turbulent medium, but only the effect
of a single compressive wave of intermediate-strength (Mach number $\sim
2$) on the already-segregated warm stable phase. Instead, here we are
considering the case of a medium with mean density $n \sim 1$
cm$^{-3}$, which is close to the ISM average density \cite{Fer01} and
lies in the unstable range, so that, even if
the two phases are segregated, the mean density remains at that
value. Moreover, rather than the effect of a single large-scale
compressive mode, we are considering a globally turbulent medium in
which the source of energy is stellar-like, having the effect of
recycling matter from the dense phase into the diffuse one. As shown in
the simulations, a situation similar to that of Hennebelle \& P\'erault
applies at the sites where the compressions are strong (near stellar
sources), except that in their case the triggering is in absolutely
linearly stable gas, while in our case the triggering is in gas stable
only to adiabatic perturbations. On the other hand, at more
remote sites, where the turbulence is weaker \cite{AV01}, the
fluctuations remain stable in general.

\begin{figure}
\begin{center}
\includegraphics[width=0.7\textwidth]{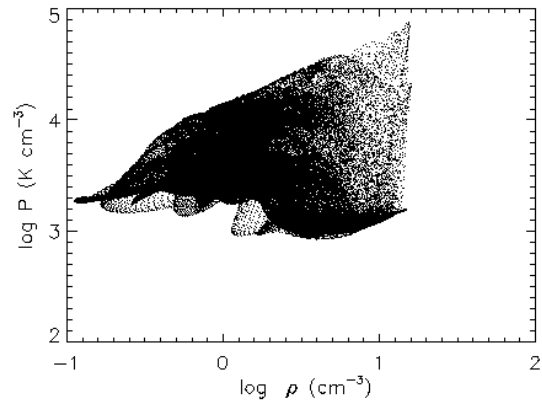}
\end{center}
\caption[]{Pressure vs.\ density for a vertical stripe of width 200 pixels
($\sim 20$ pc) in the 150-pc simulation, passing through the large
complex slightly up and to the left of the simulation center (cf. fig.\
\ref{fig:150pc_den}).}
\label{fig:P_rho_150pc}
\end{figure}

It should be emphasized that, in the context of full ISM simulations,
the regions with a reversed pressure gradient are seen to occupy a
very small fraction of the volume, with the majority of the space being
occupied by a moderately 
turbulent medium in a nearly isobaric regime, but with significant
fluctuations around it. Particularly noteworthy is the existence of
weak expanding shock waves which propagate away from the filamentary
clouds, behind which {\it both} the density and the pressure are
slightly increased with respect to the intercloud medium, similarly to
the case of the wave observed in run DEN75 in \S \ref{sec:pure_TI}. It
should be noted, however, that the origin of the outgoing shock waves
is not exactly the same in the ISM simulations as it was in run DEN75.
In the latter, the shock originates when the condensation reaches its
maximum (overshooting) density and terminates the converging motions
within it. In the ISM
simulations, it originates when star formation suddenly heats and
repressurizes the clouds. Nevertheless, the effect on the gas
surrounding the clouds is similar. The volume
between the wave front and the central clouds contains gas with a
``regular'' pressure gradient (\S \ref{sec:vel_fluc}) and temperatures
in the ``unstable'' range. However, as in the case of run DEN75, this
gas is not truly unstable, as it is out of thermal equilibrium, albeit
close to pressure equilibrium. The existence of gas with a regular
pressure gradient is seen in fig.\
\ref{fig:P_rho_150pc}, which shows the thermal pressure vs. the density
for a vertical stripe of width 200 pixels ($\sim 20$ pc) and length
equal to the box size in the 150-pc
simulation including most of the large complex slightly up and to 
the left of the simulation center. It can be seen that the pressure is
far from having a unique value, and, in particular, has a scatter of
about one order of magnitude in the ``unstable'' range, with the upper
envelope of the points having a regular dependence of pressure  on
density. This effect is even more pronounced in the simulations of Mac
Low et al.\ \cite{MBKA01}, which include supernova energy input (see
also the chapter by Mac Low, this volume).

\begin{figure}
\begin{center}
\includegraphics[width=0.5\textwidth]{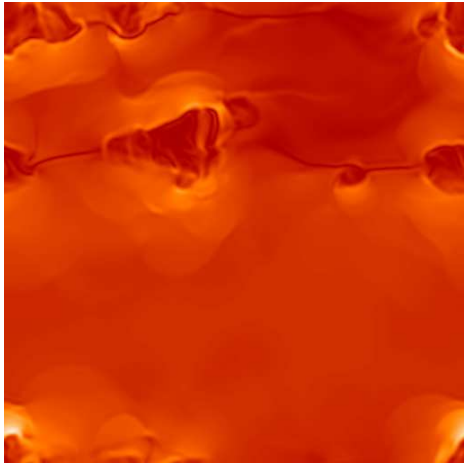}
\end{center}
\caption[]{Magnetic pressure field of the 150-pc ISM simulation. Note
that some filaments of low thermal pressure (fig.\ \ref{fig:150pc_pres})
also have low magnetic pressure, while some others have undisturbed
magnetic pressure.}
\label{fig:150pc_magn}
\end{figure}

In order to complete the description of the dynamics of the
low-thermal-pressure 
(low-$P_T$) regions, it is important to also analyze the magnetic
pressure within them. This is shown in fig.\ \ref{fig:150pc_magn}.
Interestingly, some of the low-$P_T$ filaments are also seen as
low-magnetic-pressure filaments, while some others are not. The cases
with reduced magnetic pressure can be interpreted as sites of field
reversals, which arise when the compressions amplify the (small-scale)
field fluctuations perpendicular to the direction of compression, as
observed by Passot, \VS\ \& Pouquet \cite{PVP95}. As the field reverses, 
it passes through zero. However, very few examples of
a filament seen also as a high-magnetic-pressure structure are seen,
although the magnetic pressure maxima are indeed seen to occur in the
star-forming regions. We thus conclude that in general the collapse of
the low-$P_T$ regions is ensured. This is confirmed by the fact that
those filaments systematically become sites of new star formation at
later times as the density threshold is reached. 

Also of interest is that the enhanced-density, enhanced-$P_T$ regions
surrounding the clouds in general also have an enhanced magnetic
pressure, so that these regions are in general slightly over-pressured
with respect to the global intercloud medium, and expand somewhat. The
intercloud medium is completely permeated by these traveling fronts
originating from the clouds. This is most clearly seen in animations,
which can be retrieved from our web site {\tt
http://www.astrosmo.unam.mx/$\sim$e.vazquez/turbulence\_HP}. Finally, the
magnetic field is rather uniform in the intercloud medium. More
discussion concerning the magnetic pressure in ISM simulations can be
found in the chapter by Mac Low.

\subsection{Discussion and caveats} \label{sec:discussion}

The results from the previous section suggest a dynamic ISM in which
several physical processes are occurring simultaneously besides the pure 
condensation of thermally-unstable regions, and in which, in fact, the
very nonlinear development of the latter leads to complex dynamics.

These results are not free of caveats, however. Three main limitations
of the simulations prevent the results from being definitive. First, the 
need to use artificial diffusivities, especially in the continuity
equation, to broaden discontinuities out to a few grid points,
enormously shortens the range of unstable scales, typically stabilizing
scales from the resolution limit up to 1/32 or 1/16 of the box. This
implies that the fraction of unstable gas could possibly be overestimated
because scales that could fragment in the real ISM do not in the
simulations. However, ``unstable'' structures larger than those damped
by the mass diffusion are seen to exist in the simulations as well,
suggesting that the effect is real, if perhaps not as strong.

Here it is important to emphasize that at {\it large scales} the medium
{\it is} indeed unstable, but this is manifested not in the spontaneous
condensation of the structures, but in their null resistance to
compression, since typically the turbulence velocity dispersion is
supersonic, and the structures are compressed by the turbulence before
they can spontaneously condense. At small scales, on the other hand, as
indicated by the runs in \S \ref{sec:vel_fluc}, a moderate amount of
turbulence prevents the spontaneous condensation because the
fluctuations are closer to being adiabatic than isobaric.

A second caveat is that the threshold density for star formation (SF)
used in the simulations is rather low ($n_{\rm cr}=15$ cm$^{-3}$ for the
1-kpc and the 150-pc runs, and $n_{\rm cr}=25$
cm$^{-3}$ for the 10-pc run), again in order to avoid extreme gradients
that would form upon the onset of stellar heating in very small, dense 
clouds. This implies that the clouds are not allowed to reach the
pressure-equilibrium density in the dense phase. The SF scheme thus
bypasses the process of a cloud (or cloud complex) becoming
self-gravitating to start forming stars. This presents the risk that
perhaps not enough accumulation of mass is allowed in the clouds before
they engage in SF, forcing a greater fraction of the gas to
remain in transit in the unstable range towards the clouds. We feel,
however, that the essence of the process is the fact that matter is
recycled continually among the phases and, given the time scales of the
problem, that the existence of sizeable amounts of gas always traversing 
the unstable range is inevitable to some extent. More accurate estimates 
of this fraction should be possible upon the application of
modifications to the numerical method, as decribed above.

The third caveat is the omission of supernova- (SN-)like energy imput,
which should sweep a larger fraction of the volume and induce the
formation of a hot phase, which we have neglected. Their inclusion would 
cause that a larger fraction of the volume would be swept up into
shells, forcing the conversion of diffuse gas into the dense
phase. Nevertheless, the turbulent mixing ought to be even stronger in
this case, and relatively quiescent regions must still exist, as observations
suggest. These would undergo slow condensation in the manner outlined
here.

In any case, the results presented here should only be considered as
suggestive of the existence of sizeable amounts of unstable gas in the
ISM, and more specifically designed numerical methods should be applied
in order to obtain more definitive confirmation.

\section{Summary and conclusions} \label{sec:conclusions}

In this paper we have reviewed results from various studies aiming 
at understanding the role of TI in the turbulent atomic ISM, and the
behavior of the magnetic pressure in the fully turbulent case. The
motivation has been twofold. On the one hand, the classic multi-phase
models of the 
ISM have neglected the implications of the ISM being turbulent, and it
is thus important to assess the consequences of advection
on the thermal and spatial structure of this medium. On the
other hand, observations have often suggested the presence of gas with
temperatures in the thermally unstable range, in apparent
contradiction with the multi-phase models.

We first reviewed the classic instability analysis of Field
\cite{Fie65}, emphasizing the different behavior of long- and
short-wavelength perturbations (for which the ratio $\eta$ of the
cooling [$\tc$] to the sound crossing [$\ts$] time is respectively small
and large), and of entropy and isentropic perturbations (which
trigger the condensation and the wave modes, respectively). We pointed
out that, while much study has been devoted to isobaric entropy
perturbations, real-world fluctuations in the ISM are produced
through velocity fluctuations which, in the small-scale limit, belong to 
the isentropic kind, and are therefore stable to first order at small
scales. We also briefly reviewed the magnetic case, in which the
presence of a uniform magnetic field can stabilize perturbations with
wavenumbers perpendicular to it.

We then reviewed results on the nonlinear stages of evolution of
isobaric entropy perturbations, focusing on those that have quantified
the magnitude of the speeds developed and the times required for
completing the condensation process as a function of the parameter
$\eta$. For presently-accepted values of the heating and cooling rates
\cite{Wol95}, large-scale initial perturbations ($\gtrsim 15$ pc, $\eta
\lesssim 0.2$) develop supersonic 
speeds, require times $\gtrsim 10$ Myr to complete the condensation
process, and end up with densities and pressures above the thermal
equilibrium value due to the ram pressure of the still infalling
gas. Those times are long compared with typical times between successive 
external shock passages, and star formation time scales. We thus
concluded that clouds formed from perturbations of such sizes (although
the resulting cloud has a size $\sim 1$ pc) are unlikely to exist in
thermal pressure equilibrium with their surroundings. Initial perturbations of
sizes $\lesssim 3$ pc, on the other hand, require times $\sim 4$ Myr to
complete their evolution and do not generate supersonic speeds, thus
reaching a more quiescent final state, and adhering better to the
paradigm of thermal-pressure bounded clouds at the end of their
evolution, although, by the time the 
cloud has formed, accretion is still occurring, so that the clouds are
bounded by weak accretion fronts, rather than contact
discontinuities. Furthermore, the gas still accreting is necessarily in
the unstable temperature range, although it is not in thermal
equilibrium; instead, it has a ``regular'' pressure behavior ($P$
increases with $\rho$), and thus it is not prone to further
fragmentation. 

We then described the evolution of perturbations
induced by turbulent random forcing. In this case the crossing time
entering $\eta$ should be taken as the minimum of the sound and the
turbulent ($\tu$) crossing times. Thus, small-scale ($\sim 0.3$ pc) {\it
velocity} fluctuations are quasi-isobaric at very small
amplitudes, because in this case $\tu > \tc > \ts$ so that the flow can
cool in response to the velocity perturbation. As the perturbation
amplitude is increased, so that $\tc > \ts > \tu$, the situation changes
because now the density is driven by the turbulent velocity rather than
by sound waves, and the perturbations become quasi-adiabatic in
character, {\it becoming stable}. We empirically found this to occur
roughly when the rms Mach number $\gtrsim 0.3$. Finally, however, if the 
perturbation amplitude becomes very large, then the density increment
induced by it becomes nonlinear and accelerates the cooling rate,
effectively causing $\eta <1$. In this case, velocity fluctuations
trigger the condensation mode, which is again unstable, and cause
condensations. 

Thus, we reached the important conclusion that small-scale fluctuations
behave very differently when they are entropy perturbations (caused, for 
example, by local variations in the heating or cooling rates) and when
they are adiabatic (caused by velocity fluctuations), being unstable
(and with the fastest growth rates)
in the former case, but linearly stable in the latter.

We then considered the magnetic field as an additional source of
pressure in the ISM, confirming earlier results that at low and
intermediate densities the magnetic pressure is strongly decorrelated
from density in fully turbulent cases (large field fluctuations), and
proposed an interpretation of this phenomenon in terms of the scaling of 
$B^2$ with density for the slow and fast modes of simple nonlinear MHD
waves. The decorrelation between magnetic pressure and density has
several implications, among which is that the magnetic field probably is
ineffective in supplementing thermal pressure in highly turbulent,
thermally unstable
conditions, and that it is probably inadequate to model magnetic
pressure by means of an equivalent polytropic behavior in the fully
turbulent case.

Finally, we discussed results from simulations of the ISM in more than
one dimension at large and
intermediate scales and at various resolutions. To this end, we first
performed a detailed study of the competition between numerical
diffusivities and the growth of TI, finding that even when the
diffusivities (especially the mass diffusion, which is necessary
numerically) are confined to the smallest scales on the numerical grid,
they can push the smallest unstable scale (the ``Field'' length $\lf$) to
relatively large scales in the simulations, especially for small
physical simulation sizes, because $\lf$ scales as $\lambda_K^{1/2}$, where
$\lambda_K$ is the diffusive scale. 

With this information, we discussed the fact that many ISM simulations
suggest that the basic structure does not depend sensitively on whether
TI is present, as long as there are turbulent motions driven by
stellar-like sources (that imply recycling of gas from the cold to the
warm phase), and that significant fractions of the gas mass
(15-50\%) appear to be in the unstable regime. This appears to be a
consequence of the fact that the diffuse medium is in a moderately
turbulent state, so that a) the fluctuations there have a regular
pressure gradient and b) the magnetic field is not strongly turbulent, and
therefore may cause additional stability.
Of course, when the relatively quiescent intercloud medium is hit by a
strong shock from, say, a supernova remnant, then TI can be rapidly
induced, as in the studies by Hennebelle \& P\'erault \cite{HP99,HP00}
and Koyama \& Inutsuka \cite{KI01}.

A final remark of interest is that it may be possible to determine
observationally whether the gas seen at unstable temperatures
corresponds to the out-of-thermal-equilibrium gas observed in the
simulations by either a) simultaneously determing two of its thermodynamic
variables, or b) comparing directly observed 
cooling rates (e.g., fine structure lines) with theoretical estimates 
of the heating rate (e.g., photoelectric heating) in specific
regions (C.\ Heiles, private communication). If this is confirmed, then
it would provide strong evidence 
that turbulent motions populate all regions of the thermodynamic
variable space, preventing a sharp segregation of the atomic ISM into
the stable phases of TI.

\bigskip
We have greatly benefitted from exchanges with C. Heiles,
P. Hennebelle, H. Koyama, J. Scalo and E. Zweibel. The report from an
anonymous referee prompted much improvement of the paper and led us to
the study of numerical damping of the growth rates. This work has
received partial financial support from CONACYT grant 27752-E, from the
French national program PCMI, and from the conference organizers to
E.V.-S. We have made extensive use of NASA's Astrophysics Data System
Abstract Service.

\end{document}